\documentclass[aoas, preprint]{imsart}

\RequirePackage[colorlinks,citecolor=blue,urlcolor=blue]{hyperref}
\RequirePackage{natbib}
\RequirePackage[OT1]{fontenc}
\usepackage{amsmath}
\usepackage{amsthm}
\usepackage{amssymb}
\usepackage{mathtools}
\usepackage{graphicx,psfrag,epsf}
\usepackage{enumerate}
\usepackage{natbib}
\usepackage{url} 
\usepackage[linesnumbered, ruled]{algorithm2e}
\usepackage{booktabs}
\usepackage{color}
\usepackage{array}
\usepackage{wrapfig}

\startlocaldefs

\numberwithin{equation}{section}
\theoremstyle{plain}
\newtheorem{prop}{Proposition}[section]
\DeclareMathOperator*{\argmin}{arg\,min}
\DeclareMathOperator*{\argmax}{arg\,max}
\DeclareMathOperator*{\tr}{tr}
\DeclareMathOperator*{\ptr}{ptr}
\SetKwRepeat{Do}{do}{while}

\endlocaldefs

\begin{document}

\begin{frontmatter}

\title{Common and Individual Structure of Brain Networks}
\runtitle{Common and Individual Structure of Brain Networks}


\begin{aug}
\author{\fnms{Lu} \snm{Wang}\thanksref{t1}\ead[label=e1]{rl.wang@duke.edu}}
\author{\fnms{Zhengwu} \snm{Zhang}\thanksref{t2}\ead[label=e2]{zhengwustat@gmail.com}}
\and
\author{\fnms{David} \snm{Dunson}\thanksref{t1}\ead[label=e3]{dunson@duke.edu}}

\runauthor{Wang, Zhang and Dunson}

\affiliation{Duke University\thanksmark{t1} and University of Rochester\thanksmark{t2}}
\end{aug}

\begin{abstract}

This article focuses on the problem of studying shared- and individual-specific structure in replicated networks or graph-valued data.  In particular, the observed data consist of $n$ graphs, $G_i, i=1,\ldots,n$, with each graph consisting of a collection of edges between $V$ nodes. In brain connectomics, the graph for an individual corresponds to a set of interconnections among brain regions.  Such data can be organized as a $V \times V$ binary adjacency matrix $A_i$ for each $i$, with ones indicating an edge between a pair of nodes and zeros indicating no edge. When nodes have a shared meaning across replicates $i=1,\ldots,n$, it becomes of substantial interest to study similarities and differences in the adjacency matrices. To address this problem, we propose a method to estimate a common structure and low-dimensional individual-specific deviations from replicated networks. The proposed Multiple GRAph Factorization (M-GRAF) model relies on a logistic regression mapping combined with a hierarchical eigenvalue decomposition. We develop an efficient algorithm for estimation and study basic properties of our approach.  Simulation studies show excellent operating characteristics and we apply the method to human brain connectomics data. 

\end{abstract}



\end{frontmatter}

\section{Introduction}
\label{sec:intro}

Binary undirected networks, encoding the presence or absence of connections between pairs of nodes, have wide applications in biology and social science \citep{girvan2002community}. While most available procedures focus on modeling a single network, we consider the case where a network over a common set of nodes is measured for each individual under study, leading to multiple network observations. One particular example is structural or functional brain networks, with the brain parcellated into a fixed number of regions. Multimodal magnetic resonance imaging (MRI) scans, together with advanced image processing tools, can give us a connectivity pattern of the brain represented by an undirected binary network \citep{Zhang2017HCP}.  Such networks from multiple subjects typically share a common structure while exhibiting their own features. 

In this context, it becomes of particular interest to study similarities and differences in human brain networks. Shared connectivity patterns provide important insights into evolutionarily-conserved structures in the human brain. For example, pairs of brain regions that have very high or very low probabilities of connection for essentially all individuals. Individual-specific structure of the brain network may help to predict and add to mechanistic understanding of causes of variation in human cognitive traits and behaviors. \citet{lock2013joint} proposed a useful tool to separate joint and individual variation for multiple datasets associated with a common set of objects. However, their method was not designed for network-valued data. There is a strong need for new statistical methods that identify and separate the common and individual structure for multiple replicated networks.

\begin{figure}[!htb]
\includegraphics[width=\textwidth]{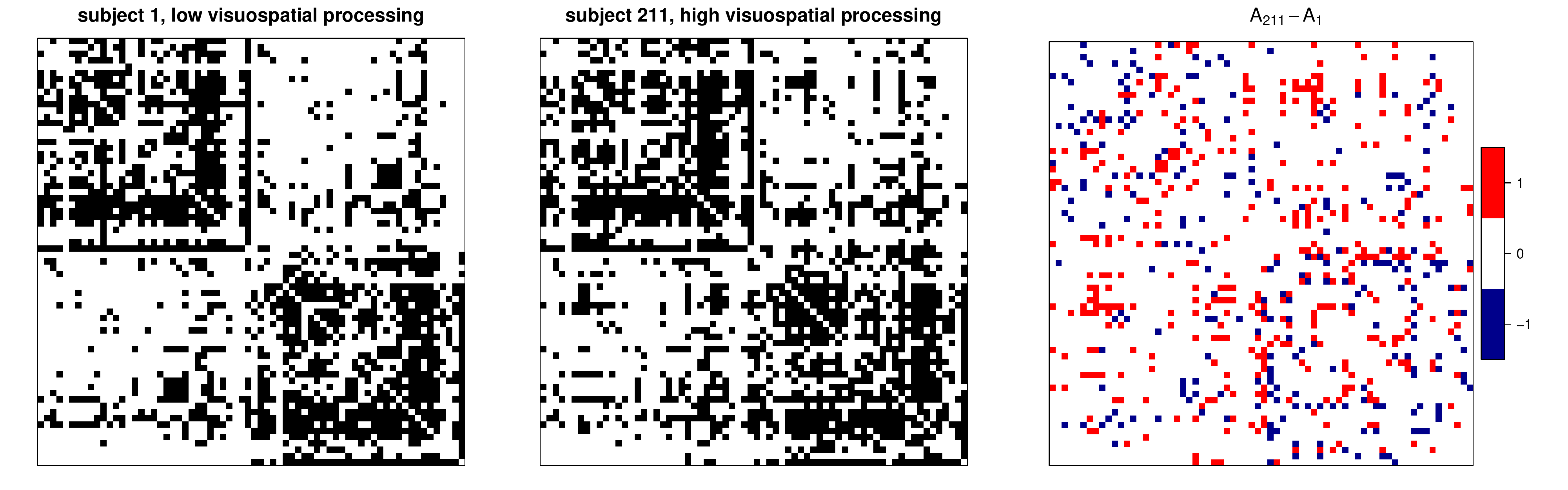}
\caption{Adjacency matrices of two structural brain networks in the HCP data (left and middle) and a heatmap of their differences (right). \label{fig:adj-visuo}}
\end{figure}

The focus of this article is on extracting common and low-dimensional individual-specific structure from replicated binary networks. In structural brain connectivity applications, providing the main motivation of this article, the individual-specific components reflect distinct characteristics of that individual's brain structure which may relate to her traits.  We focus on data from the Human Connectome Project (HCP) \citep{van2012human} (\url{www.humanconnectomeproject.org/}), which contains rich brain imaging data along with a range of cognitive, motor, sensory and emotional traits \citep{barch2013function}. Figure \ref{fig:adj-visuo} displays two binary structural brain networks we extracted from two HCP subjects and the difference of their adjacency matrices.
The left panel shows the network for an individual with a low visuospatial processing score, while the middle panel shows one with a high score.  Potentially, the individual difference, e.g. the cross-hemisphere connectivity, may predict a range of traits, such as cognitive, motor, and sensory abilities.  

There is a large literature on statistical modeling of binary networks \citep{goldenberg2010survey}. For example, exponential family random graph models (ERGMs) assume the probability of observing a graph is determined by a vector of graph statistics, such as the total number of edges, the degrees of the vertices and so on. However, since the mappings from graphs to features are often many to one mappings, one drawback of the ERGM is that simply relying on some summary features of a graph can not represent complex structures of networks. Latent space models \citep{hoff2002latent}, however, are more flexible at characterizing the distribution of graphs because they can effectively model each edge probability, while still maintaining rich types of dependence structure in the graph. 

A variety of latent space models have been developed  \citep{hoff2008eigenmodel, tang2016law, durante2017nonparametric}, which are appealing in defining rich types of network structure while achieving dimensionality reduction by embedding each node in a low dimensional latent space.  Edges are typically assumed to be conditionally independent given the latent positions of nodes. The edge probabilities are described as functions of distance or (weighted) inner products of node-specific latent vectors with a {\it logit} or {\it probit} link. Bayesian inference is often employed, but
substantial computational problems can arise for large multiple-network data. 

Considering a single network, efficient algorithms have been developed for estimating its low dimensional latent structure. \citet{sussman2012consistent} estimate nodes' latent positions from a low rank approximation to the adjacency matrix for a random dot product graph (RDPG) with identity link. Though \citet{sussman2012consistent} proved the consistency of assigning nodes to blocks by clustering over their latent vectors, the dot product of the estimated latent positions may not be valid probabilities. \citet{o2015clustering} proposed to do node clustering on an RDPG with a logistic link to address this problem. They provided an efficient algorithm for maximum likelihood inference of nodes' latent positions which contains a spectral decomposition on the mean-centered adjacency matrix and a logistic regression with positive constraint on the coefficients. 

There is a literature on analysis methods for data consisting of a set of networks that share a common vertex set. For multi-layer or multi-view graphs, vertices correspond to entities and different graph layers capture different types of relationships among the entities \citep{dong2014clustering}. 
Linked Matrix Factorization (LMF) \citep{tang2009clustering} approximates each graph by a graph-specific factor and a factor matrix common to all graphs. The goal is to cluster vertices into communities, and LMF focuses on merging the information from multiple graphs instead of characterizing unique structure for each graph. Other relevant methods include principal component analysis (PCA) and tensor decomposition \citep{tucker1966some, kolda2009tensor}.  Usual PCA requires {\em flattening} of the data, which destroys the network structure, while tensor methods that concatenate the adjacency matrices together might be more appropriate \citep{zhang2018relationships}. None of these approaches directly addresses our problem of interest.

We develop a promising framework for studying the common and low dimensional individual structure of multiple binary networks with similar patterns. Our approach provides a data generating process for each graph, which shows how the low dimensional structure drives the high dimensional networks. Specifically, the logit of the edge-probability matrix for each network is decomposed into the sum of a common term and a low-rank individual-specific deviation. Based on the idea of an unrestricted eigen-decomposition (no positive constraints on eigenvalues), our model is able to capture complex network patterns, such as hubs \citep{hoff2008eigenmodel}, better than latent distance or latent inner-product models. A novel algorithm inspired by \citet{o2015clustering} is proposed for efficiently estimating the model.

The rest of the paper is organized as follows. The model and algorithm together with two variants are proposed in Section \ref{sec:meth}. Section \ref{sec:exp} contains simulation studies demonstrating the computational performance of our algorithm and basic properties of parameter estimates. Applications to scan-rescan brain network data and the HCP data are reported in Section \ref{sec:app} and Section \ref{sec:conc} concludes.

\section{Methodology}
\label{sec:meth}
We focus on undirected binary networks with a common node set and no self-loops. Let $A_{1},\dots,A_{n}$ be the corresponding adjacency matrices of these networks. Each $A_{i}$ is a $V\times V$ symmetric matrix with $A_{i[vu]}=1$ if node $u$ and $v$ are connected in network $i$ and $A_{i[vu]}=0$ otherwise. 

\subsection{M-GRAF Model}
\label{subsec:Model}
We take the conditional independence approach of latent space models by assuming for each pair of nodes $(u,v)$ in network $A_{i}$, an edge is drawn independently from a Bernoulli distribution given the corresponding edge probability:
\begin{equation}
A_{i[uv]}\mid\Pi_{i[uv]}\stackrel{\mbox{ind}}{\sim}\mbox{Bernoulli}(\Pi_{i[uv]}),\ u>v;u,v\in\{1,2,\dots,V\},\label{eq:model1}
\end{equation}
where $\Pi_{i}$ denotes the $V\times V$ symmetric edge probability matrix corresponding to network $i$, $i=1,\dots,n$. 

\begin{figure}[!htb]
\includegraphics[width=\textwidth]{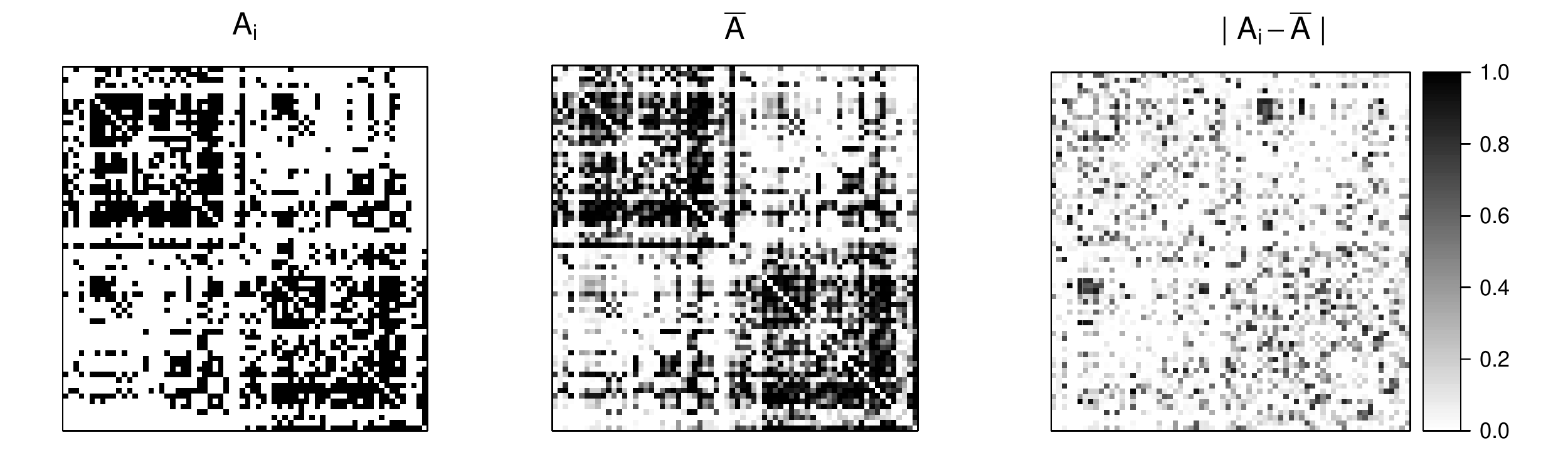}
\caption{Left: the adjacency matrix $A_{i}$ of a subject's structural brain network in the HCP data. Middle: average of the adjacency matrices $\bar{A}=\sum_{i=1}^{n}A_{i}/n$. Right: absolute value of $(A_{i}-\bar{A})$. \label{fig: adjacency}}
\end{figure}

In our exploratory analyses of brain network data, we observe that brain structural networks generally share some common connectivity patterns such as hemisphere modularity, as shown in Figure \ref{fig: adjacency}. In addition, the deviation of individual networks from the average tends to be much sparser, with many entries in the deviation matrix $\left|A_{i}-\bar{A}\right|$ of small magnitude (shown in the right most panel of Figure \ref{fig: adjacency}). We expect that these deviations can be accurately approximated as low rank. 

Therefore, we assume the logit of each probability matrix $\Pi_{i}$ can be decomposed into two parts: a symmetric matrix $Z\in\mathbb{R}^{V\times V}$ shared by all networks representing the baseline log odds for each connection and a symmetric low rank matrix $D_i \in\mathbb{R}^{V\times V}$ representing the deviation of unit $i$ from the baseline:
\begin{equation}
\mbox{logit}(\Pi_{i})=Z+D_{i},\ i=1,\dots,n.\label{eq:edge_prob_fact}
\end{equation}
Suppose $D_{i}$ has rank $K$,  typically with $K\ll V$. Taking an eigenvalue decomposition of $D_{i}$,
\begin{equation}
D_{i}=Q_{i}\Lambda_{i}Q_{i}^{\top}, \label{eq:eigen-decom}
\end{equation}
where $Q_{i}\in\mathbb{R}^{V\times K}$ satisfies $Q_{i}^{\top}Q_{i}=I_{K}$ and $\Lambda_{i}=\mbox{diag}(\lambda_{i1},\dots,\lambda_{iK})$ is a $K\times K$ diagonal matrix.

Equations \eqref{eq:edge_prob_fact} - \eqref{eq:eigen-decom} imply that the individual elements of $\Pi_i$ can be expressed as:
\begin{multline}
\mbox{logit}(\Pi_{i[uv]})=Z_{uv}+\sum_{k=1}^{K}\lambda_{ik}Q_{i[uk]}Q_{i[vk]}, \\ \mbox{for } \ u\neq v,\,u,v\in\{1,\dots,V\},i=1,\dots,n.\label{eq:logistic regression}
\end{multline}
$Z_{uv}$ in \eqref{eq:logistic regression} represents the baseline log odds for the node pair $u$, $v$ across all networks. Interpretation of the rest of \eqref{eq:logistic regression} is similar to that of the eigenmodel in \citet{hoff2008eigenmodel} where the relationship between two nodes is represented as a weighted dot product of node-specific latent vectors. For each network $i$, $Q_{i[uk]}$ can be interpreted as node $u$'s value of some unobserved characteristic $k$ or latent coordinate along axis $k$. $\lambda_{ik}$ is the scaling parameter of latent axis $k$. The magnitude of $\lambda_{ik}$ controls the impact of axis $k$ in determining the edge probabilities of network $i$, while the sign of $\lambda_{ik}$ determines whether similar values of $Q_{i[uk]}$ and $Q_{i[vk]}$ would contribute positively or negatively to the connection probability between node $u$ and $v$. 

This model based on the idea of eigenvalue decomposition is flexible at characterizing a wide array of patterns in network data  \citep{hoff2008eigenmodel}, including transitivity and hubs. Transitivity describes the phenomenon that a friend of a friend is a friend, which is well represented by a latent distance model or RDPG but is poorly characterized by a stochastic block model. A hub refers to a center node that connects to many other nodes while these nodes do not connect to each other. Such structure could be described by a stochastic block model with a large number of groups. However, latent distance models or RDPGs often perform poorly or require high dimension of latent characteristics to capture the pattern of hubs. As \citet{hoff2008eigenmodel} pointed out, the flexibility of the eigenmodel is ``due to the fact that it provides an unrestricted low rank approximation to the adjacency matrix'' and is therefore able to represent more complicated patterns than the other three latent space models.

\subsection{Inference of $Q_{1},\dots,Q_{n}$}
\label{subsec:MLE_Q}

For estimation of the model \eqref{eq:model1} - \eqref{eq:eigen-decom}, we first simplify the joint log-likelihood of the $n$ network observations \\$A_{1},\dots,A_{n}$ as
\begin{align}
\log L(A_{1},\dots,A_{n}\mid Z,D_{1},\dots,D_{n})=\qquad\qquad\qquad\qquad\qquad\qquad\qquad \nonumber\\
\sum_{i=1}^{n}\sum_{u=1}^{V}\sum_{v<u}\left[A_{i[uv]}\log(\Pi_{i[uv]})+(1-A_{i[uv]})\log(1-\Pi_{i[uv]})\right] \nonumber \\
=\sum_{i=1}^{n}\sum_{u=1}^{V}\sum_{v<u}\left[A_{i[uv]}\log\left(\dfrac{\Pi_{i[uv]}}{1-\Pi_{i[uv]}}\right)+\log(1-\Pi_{i[uv]})\right]\quad \nonumber\\
=\sum_{i=1}^{n}\sum_{u=1}^{V}\sum_{v<u}\left[A_{i[uv]}(Z_{uv}+D_{i[uv]})+\log(1-\Pi_{i[uv]})\right].\qquad\ 
\label{eq: joint-loglikelihood}
\end{align}
\begin{prop} \label{prop:trace}
Assuming that the common structure $Z$ is given,
\[
\underset{D_{1},\dots,D_{n}}{\argmax\ }\log L(A_{1},\dots,A_{n}\mid Z,D_{1},\dots,D_{n})=\underset{D_{1},\dots,D_{n}}{\argmax\ }\sum_{i=1}^{n}\dfrac{1}{2}\tr \left([A_{i}-\pi(Z)]D_{i}\right),
\]
where $\tr(\cdot)$ is the matrix trace, $\pi(\cdot)$ is the logistic function and $\pi(Z)$ denotes applying $\pi(\cdot)$ to each entry in matrix $Z$. The diagonal elements of $\pi(Z)$ and  $A_{i}$ are set to $0$.
\end{prop}

The proof of Proposition \ref{prop:trace} can be found in Appendix \ref{app:prop1}. From the form of the joint log-likelihood \eqref{eq: joint-loglikelihood}, it is clear the $D_{i}$s can be estimated independently for $i=1,\ldots,n$ conditionally on $Z$.  According to Proposition \ref{prop:trace}, $\tr \left([A_{i}-\pi(Z)]D_{i}\right)$ is a good surrogate function of the log-likelihood $\log L(A_{i} \mid Z,D_{i})$, which is easier to maximize since it is linear in $D_i$. Hence given $Z$ and recalling the low rank assumption on $D_{i}$, we solve the following optimization \eqref{eq:optim_D} to estimate $D_i$:
\begin{eqnarray}
\underset{D_{i}}{\max} & \tr\left([A_{i}-\pi(Z)]D_{i}\right)\label{eq:optim_D}\\
\mbox{s.t.} & \mbox{rank}(D_{i})=K.\nonumber 
\end{eqnarray}
Plugging in the eigen-decomposition \eqref{eq:eigen-decom} of $D_{i}$ into the target function of \eqref{eq:optim_D}, we have
\begin{eqnarray*}
\tr \left([A_{i}-\pi(Z)]D_{i}\right) & = & \tr \left([A_{i}-\pi(Z)]Q_{i}\Lambda_{i}Q_{i}^{\top}\right)\\
 & = & \tr \left(Q_{i}^{\top}[A_{i}-\pi(Z)]Q_{i}\Lambda_{i}\right)\\
 & = & \sum_{k=1}^{K}\lambda_{ik}Q_{i[\cdot k]}^{\top}[A_{i}-\pi(Z)]Q_{i[\cdot k]}
\end{eqnarray*}
where $Q_{i[\cdot k]}$ denotes the $k$-th column of $Q_{i}$. Then we obtain the following equivalent optimization to \eqref{eq:optim_D}.
\begin{eqnarray}
\underset{Q_{i},\Lambda_{i}}{\max} & \sum_{k=1}^{K}\lambda_{ik}Q_{i[\cdot k]}^{\top}[A_{i}-\pi(Z)]Q_{i[\cdot k]}\label{eq:optim_Q}\\
\mbox{s.t.} & Q_{i}^{\top}Q_{i}=I_{K},\ Q_{i}\in\mathbb{R}^{V\times K}.\nonumber 
\end{eqnarray}
Suppose the diagonal entries of $\Lambda_{i}$ are sorted decreasingly so that $\lambda_{i1}\geq\dots\geq\lambda_{ik}>0>\lambda_{i,k+1}\dots\geq\lambda_{iK}$. Then the optimal $Q_i$ in \eqref{eq:optim_Q} can be solved according to the following Proposition \ref{prop:prop2}.

\begin{prop}
\label{prop:prop2}
Let $B$ be a $V\times V$ symmetric real matrix. Suppose the eigenvalues of $B$ are $\sigma_{1}(B)\geq\cdots\geq\sigma_{V}(B)$ and the corresponding orthonormal eigenvectors are $\boldsymbol{q}_{1},\dots,\boldsymbol{q}_{V}$. For any $k\in\{1,\ldots,V\}$, given $k$ positive real numbers $c_{1}\geq\cdots\geq c_{k}>0$, and for any orthonormal set $\{\boldsymbol{u}_{1},\dots,\boldsymbol{u}_{k}\}$ in $\mathbb{R}^{V}$, one has


\begin{eqnarray}
\underset{\boldsymbol{u}_{1},\dots,\boldsymbol{u}_{k}}{\max}\sum_{j=1}^{k}c_{j}\boldsymbol{u}_{j}^{\top}B\boldsymbol{u}_{j} & = & 
 c_{1}\sigma_{1}(B)+\cdots+c_{k}\sigma_{k}(B)
 \label{eq:prop_sup}
\end{eqnarray}
and
\begin{eqnarray}
\underset{\boldsymbol{u}_{1},\dots,\boldsymbol{u}_{k}}{\min}\sum_{j=1}^{k}c_{j}\boldsymbol{u}_{j}^{\top}B\boldsymbol{u}_{j} & = & c_{1}\sigma_{V}(B)+\cdots c_{k}\sigma_{V-k+1}(B).\label{eq:prop_inf}
\end{eqnarray}
Therefore an optimal solution to \eqref{eq:prop_sup} is $\{\boldsymbol{q}_{1},\dots,\boldsymbol{q}_{k}\}$ and an optimal solution to \eqref{eq:prop_inf} is $\{\boldsymbol{q}_{V},\dots,\boldsymbol{q}_{V-k+1}\}$.
\end{prop}

The proof is in Appendix \ref{app:prop2}. 
Let $\boldsymbol{q}_{1}^{(i)},\dots,\boldsymbol{q}_{k}^{(i)}$ be the first $k$ eigenvectors of $A_{i}-\pi(Z)$ corresponding to the largest eigenvalues, and $\boldsymbol{q}_{V-K+k+1}^{(i)},\dots,\boldsymbol{q}_{V}^{(i)}$ the last $(K-k)$ eigenvectors of $A_{i}-\pi(Z)$ corresponding to the smallest eigenvalues. Then according to Proposition \ref{prop:prop2}, an optimal solution $Q_{i}$ to \eqref{eq:optim_Q} is $Q_{i}=(\boldsymbol{q}_{1}^{(i)},\dots,\boldsymbol{q}_{k}^{(i)},\boldsymbol{q}_{V-K+k+1}^{(i)},\dots,\boldsymbol{q}_{V}^{(i)})$.

\subsection{Logistic Regression for $Z$ and $\{\lambda_{ik}\}$}
\label{subsec:Bayesian-Logistic-Regression}

Once $\{Q_{i}:i=1,\dots,n\}$ is estimated, it remains only to estimate the parameters $\{\lambda_{ik}:k=1\dots,K;i=1\dots,n\}$ and $Z$. Note that $\lambda_{ik}$'s and entries of $Z$ are linear in the logistic link function \eqref{eq:logistic regression}. Therefore the MLE of $\{\lambda_{ik}\}$ and $Z$ given $\{Q_{i}\}$ can be solved by logistic regression of the lower triangular entries of $\{A_{i}:i=1,\dots,n\}$ on the corresponding entries of $\{Q_{i[\cdot k]}Q_{i[\cdot k]}^{\top}:k=1,\dots,K;i=1,\dots,n\}$. Let $\mathcal{L}(\cdot)$ be a function mapping the lower triangular entries of a $V\times V$ matrix into a $V(V-1)/2\times1$ long vector, let $\boldsymbol{\pi}_{i}=\mathcal{L}(\Pi^{(i)})=(\pi_{i1},\dots,\pi_{iL})^{\top}$, where $L=V(V-1)/2$, let $\boldsymbol{z}={\cal L}(Z)=(z_{1},\dots,z_{L})^{\top}$, and let $M_{i}$ be a $L\times K$ matrix with each column being $M_{i[\cdot k]}={\cal L}(Q_{i[\cdot k]}Q_{i[\cdot k]}^{\top})$ for $k=1,\dots,K$. Then \eqref{eq:logistic regression} can be written as
\begin{eqnarray}
\mbox{logit}(\pi_{il}) & =z_{l}+\sum_{k=1}^{K}\lambda_{ik}M_{i[lk]}, & l=1,\dots,L;i=1,\dots,n.\label{eq:logistic_vector}
\end{eqnarray}

However, as $K$ increases, overfitting could cause a serious separation issue in the logistic regression \eqref{eq:logistic_vector}, where the binary outcomes can be almost perfectly predicted by a linear combination of predictors. The separation issue is well known to cause nonidentifiability of logistic regression coefficients with the MLE being $\pm \infty$. A solution to this problem is to place a penalty or prior on the coefficients. Penalized likelihood estimation proposed by \citet{firth1993bias} is equivalent to the use of Jeffreys invariant prior. The Newton-Raphson algorithm by \citet{heinze2006penalize} based on Firth's method was very slow even for a small synthetic dataset in our simulation. \citet{gelman2008glm_prior} propose independent Cauchy priors with center 0 and scale 2.5 for each of the  logistic regression coefficients as a weakly informative default.  However, such Cauchy priors have very heavy tails and often do not have good performance in sparse data settings with separation issues in our experience. Hence, we instead recommend the following weakly informative Gaussian prior distributions:
\begin{align}
Z_{uv}\sim & N(0,10^{2} / \gamma),\ u>v,\,u,v\in\{1,\dots,V\}\label{eq:prior_Z}\\
\lambda_{ik}\sim & N\left(0,\dfrac{2.5^{2}}{\gamma \cdot (2\cdot\mathtt{sd}_{ik})^{2}}\right),\ k=1,\dots K;i=1,\dots,n, \label{eq:prior_lambda}
\end{align}
where $\gamma$ is a prior precision factor, $\mathtt{sd}_{ik}$ is the standard deviation (sd) of $M_{i[\cdot k]}$, and the factor $1/(2\cdot\mathtt{sd}_{ik})^{2}$ in \eqref{eq:prior_lambda} is equivalent to standardizing the predictors to have sd of 0.5 as suggested by \citet{gelman2008glm_prior}. 

The Gaussian prior is equivalent to $L_2$ regularization or a ridge penalty for generalized linear models. Hence, we could compute maximum-a-posteriori (MAP) estimates for  $Z$ and $\{\lambda_{ik}\}$ with the \texttt{glmnet} function in R. The algorithm implemented in \texttt{glmnet} uses cyclical coordinate descent \citep{Friedman2010aa} and can handle large problems efficiently in our experience. $\gamma$ is selected through cross validation. 

\subsection{CISE Algorithm}

Based on the derivations above, we develop a CISE (common and individual structure explained) algorithm for estimating the M-GRAF model \eqref{eq:model1} - \eqref{eq:eigen-decom}. CISE is essentially a block coordinate descent algorithm and Algorithm \ref{alg:Joint-inference} presents the details.  
\begin{algorithm}[!htb]
\KwIn{Adjacency matrices $A_{1},\dots,A_{n}$ of size $V \times V$, low rank $K$, tolerance $\epsilon \in \mathcal{R}_+$.}
\KwOut{Estimates of $\{Q_{i}:i=1,\dots,n\}$, $Z$ and $\{\lambda_{ik}:k=1\dots,K;i=1\dots,n\}$. } 
Initialize $\hat{\pi}(Z)=\sum_{i=1}^{n}A_i / n$ \; 
Initialize each $\hat{Q}_{i}$ to be the $K$ eigenvectors of $A_{i}-\hat{\pi}(Z)$ corresponding to the largest eigenvalues in magnitude.

\Do{percent change of joint log-likelihood \eqref{eq: joint-loglikelihood} $\geq \epsilon$ }{
(I) Perform $L_2$-penalized logistic regression \eqref{eq:logistic_vector} - \eqref{eq:prior_lambda}  to obtain the MAP estimates of $\{Z_{uv}\}$ and $\{\lambda_{ik}\}$. Time complexity of this step is $O(nV^2K)$, according to \citet{minka2003comparison}. 

(II) For each $i$, let $k_{i}$ be the number of positive values in $\lambda_{i,1:K}$ \; 
\qquad Compute the first $k_{i}$ eigenvectors of $A_i -\pi(Z)$, $\boldsymbol{q}_{1}^{(i)},\dots,\boldsymbol{q}_{k_{i}}^{(i)}$,  and the last \\
\qquad $(K-k_{i})$ eigenvectors, $\boldsymbol{q}_{V-K+k_{i}+1}^{(i)},\dots,\boldsymbol{q}_{V}^{(i)}$ (with sorted eigenvalues).\\
\qquad Let $\hat{Q}_i =(\boldsymbol{q}_{1}^{(i)},\dots,\boldsymbol{q}_{k_{i}}^{(i)},\boldsymbol{q}_{V-K+k_{i}+1}^{(i)},\dots,\boldsymbol{q}_{V}^{(i)})$. Time complexity of this \\
\qquad partial eigen-decomposition is $O(V^2 K)$ or less \citep{woolfe2008fast}.
}
\caption{Common and individual structure explained (CISE) for multiple binary networks. \label{alg:Joint-inference}}
\end{algorithm}

\subsection{Distance-based Classification of Networks}
\label{subsec:classification}
In many applications, in addition to the network variable $A_i$ there may be a class label $l_{i}\in\{1,2,\dots,m\}$ associated with each subject $i$ in the dataset, such as high IQ or low IQ, healthy or with Alzheimer's disease. People may want to predict the class membership for a new unlabeled subject based on her brain connectivity. After estimating the low-rank components $\{Q_i,\Lambda_i\}$, representing individual-specific features of a subject's network data, classification can proceed via a simple distance-based procedure. We define the following distance measure between subject $i$ and $j$, which avoids misalignment and rotation issues of eigenvectors across subjects:
\[
d(i,j)\coloneqq\left\Vert D_{i}-D_{j}\right\Vert _{F}=\left\Vert Q_{i}\Lambda_{i}Q_{i}^{\top}-Q_{j}\Lambda_{j}Q_{j}^{\top}\right\Vert _{F},
\]
where $\left\Vert \cdot\right\Vert _{F}$ denotes the Frobenius norm. Since $Q_{i}$ and $Q_{j}$ lie on the Stiefel manifold ${\cal S}_{K,V}=\{X\in\mathbb{R}^{V\times K}:X^{\top}X=I_{K}\}$, we can further simplify this distance metric as
\begin{eqnarray*}
d^{2}(i,j) & = & \tr \left[(Q_{i}\Lambda_{i}Q_{i}^{\top}-Q_{j}\Lambda_{j}Q_{j}^{\top})^{\top}(Q_{i}\Lambda_{i}Q_{i}^{\top}-Q_{j}\Lambda_{j}Q_{j}^{\top})\right]\\
 & = & \tr \left(\Lambda_{i}^{2}\right)+\tr \left(\Lambda_{j}^{2}\right)-2\tr (\Lambda_{i}Q_{i}^{\top}Q_{j}\Lambda_{j}Q_{j}^{\top}Q_{i})
\end{eqnarray*}
so that we only need to compute traces of several small $K \times K$ matrices instead of the large $V \times V$ matrices. For a new unlabeled subject $i^{\star}$, the proximity measure between $i^{\star}$ and a class $c$ is defined as the average distance from $i^{\star}$ to all the subjects in the class $c$. Subject $i^{\star}$ is then allocated to the class with the minimum proximity.


\subsection{Variants}

The model described in Section \ref{subsec:Model} is very flexible, since for each subject, we have $\{Q_i, \Lambda_i\}$ to represent its individual structure. This model can be modified to further reduce the number of parameters in two different settings so as to accommodate different degrees of heterogeneity in the data. 

\subsubsection{Variant 1: $D_i = Q_i \Lambda Q_i^\top$}

In this case, $\Lambda_i$'s are assumed to be the same over all networks so that the number of unknown coefficients in $\{ \Lambda_i\}$ declines from $nK$ to $K$. This model implies that the scaling parameters controlling the impacts of the latent axes are equal for all networks (as discussed in Section \ref{subsec:Model}). In this case, the estimation of $Q_{i^{\star}}$ for a new network $i^{\star}$ becomes quite efficient once $Z$ and $\Lambda$ have been estimated from the training set of networks. Suppose the diagonal entries of $\hat{\Lambda}$ are sorted decreasingly: $\hat{\lambda}_{1}\geq\dots\geq\hat{\lambda}_{k}>0>\hat{\lambda}_{k+1}\dots\geq\hat{\lambda}_{K}$, $\hat{Q}_{i^{\star}}$ therefore consists of the first $k$ and the last $(K-k)$ eigenvectors of $A_{i^{\star}}-\pi(\widehat{Z})$. This variant provides competitive goodness-of-fit to the brain network data compared with the more flexible model $D_i = Q_i \Lambda_i Q_i^\top$ as shown in the applications. 

Only a small modification to Algorithm \ref{alg:Joint-inference} is needed for estimation of $\Lambda$. Again we choose a weakly informative prior for $\lambda_{k}$'s:
\begin{equation}
\lambda_{k}\sim N\left(0,\dfrac{2.5^{2}}{\gamma \cdot (2\cdot\mathtt{sd}_{k})^{2}}\right),\ k=1,\dots K, \label{eq:prior_lambda2}
\end{equation}
where $\mathtt{sd}_{k}$ is the standard deviation of $(M_{i[\cdot k]}^{\top},\dots,M_{n[\cdot k]}^{\top})^{\top}$. As in \eqref{eq:prior_lambda}, the factor $1/(2\cdot\mbox{sd}_{k})^{2}$  in \eqref{eq:prior_lambda2} is equivalent to standardizing the predictor to have sd of 0.5; the numerator 2.5 is the suggested scale of the Cauchy prior by \citet{gelman2008glm_prior}; $\gamma$ adds flexibility to the shrinkage of this prior, which is often tuned by cross validation in practice. Then the MAP estimates for $Z$ and  $\Lambda$ can be obtained via a $L_2$-penalized logistic regression. 

\subsubsection{Variant 2: $D_{i}=Q\Lambda_{i}Q^{\top}$}

Alternatively, we might do a joint embedding by restricting $Q_{i}$'s to be the same. Then the individual structure of each network is represented by a linear combination of $K$ common rank-one matrices and a $K\times1$ loading vector $\lambda_{i,1:K}$, which greatly reduces dimensionality. In this joint embedding setting, we could still follow an iterative algorithm to do inference on the parameters. $Z$ and $\{\Lambda_{i}:i=1,\dots,n\}$ can be estimated from a logistic regression with ridge penalty as discussed in Section \ref{subsec:Bayesian-Logistic-Regression} with $M_{i}$ replaced by $M={\cal L}(QQ^{\top})$. The challenge lies in estimating $Q$ given $Z$ and $\{\Lambda_{i}\}$.

Similar to the previous cases, given $Z$ and $\{\Lambda_{i}\}$, $Q$ can be estimated from the following optimization
\begin{eqnarray}
\underset{Q\in\mathbb{R}^{V\times K}}{\max} & \sum_{i=1}^{n}\tr \left([A_{i}-\pi(Z)]D_{i}\right)\label{eq:same Q}\\
\mbox{s.t.} & D_{i}=Q\Lambda_{i}Q^{\top},\nonumber \\
 & Q^{\top}Q=I_{K}.\nonumber 
\end{eqnarray}
Plugging in $D_{i}=Q\Lambda_{i}Q^{\top}$ into the target function of \eqref{eq:same Q}, we have

\begin{align*}
\sum_{i=1}^{n}tr\left([A_{i}-\pi(Z)]D_{i}\right)=\sum_{i=1}^{n}tr\left([A_{i}-\pi(Z)]Q\Lambda_{i}Q^{\top}\right)\qquad\qquad\qquad\\ 
=\sum_{i=1}^{n}tr\left(Q^{\top}[A_{i}-\pi(Z)]Q\Lambda_{i}\right)=\sum_{i=1}^{n}\sum_{k=1}^{K}\lambda_{ik}\boldsymbol{q}_{k}^{\top}[A_{i}-\pi(Z)]\boldsymbol{q}_{k}\\
=\sum_{k=1}^{K}\boldsymbol{q}_{k}^{\top}\left\{ \sum_{i=1}^{n}\lambda_{ik}[A_{i}-\pi(Z)]\right\} \boldsymbol{q}_{k}\qquad\qquad\qquad\qquad\qquad\quad\ 
\end{align*} 
where $\boldsymbol{q}_{k}$ is the $k$th column of $Q$. Define $W_{k}\coloneqq\sum_{i=1}^{n}\lambda_{ik}[A_{i}-\pi(Z)],\ k=1,\dots,K.$ 
Then the optimization \eqref{eq:same Q} can be written as
\begin{eqnarray}
\underset{\boldsymbol{q}_{1},\dots\boldsymbol{q}_{K}}{\mbox{max}} & \sum_{k=1}^{K}\boldsymbol{q}_{k}^{\top}W_{k}\boldsymbol{q}_{k}\label{eq:simplified Q}\\
\mbox{s.t.} & \boldsymbol{q}_{k}^{\top}\boldsymbol{q}_{k}=1, & \boldsymbol{q}_{k}^{\top}\boldsymbol{q}_{j}=0\ (k\neq j)\nonumber 
\end{eqnarray}
Let $\mbox{evec}_{1}(W)$ denote the first eigenvector (unit length) of $W$ corresponding to the largest eigenvalue. If $\mbox{evec}_{1}(W_{1}),\dots,\mbox{evec}_{1}(W_{K})$ are close to $K$ orthonormal vectors, we will obtain a global maxima for (\ref{eq:simplified Q}), otherwise, we can only get a local maxima due to the fact that the optimization is non-convex and there is no closed form solution available. A greedy algorithm is developed to solve (\ref{eq:simplified Q}), and the details are presented in Appendix \ref{inference:jem}.

\section{Simulation Studies}
\label{sec:exp}

In this section, we conduct a number of simulation experiments to evaluate the efficiency of CISE algorithm. We also assess the performance of M-GRAF model in inference on the common and individual-specific components of variability in synthetic networks.
CISE algorithm is implemented in both R and Matlab and all the numerical experiments are conducted in a machine with 8 Intel Core i7 3.4 GHz processor and 16 GB of RAM. The Matlab and R codes are publicly available in Github  (see \ref{suppA} for the link). The algorithm is also implemented in the R package \texttt{CISE} available on CRAN.

\subsection{Computational Performance}

\begin{figure}[!htb]
\includegraphics[width=\textwidth]{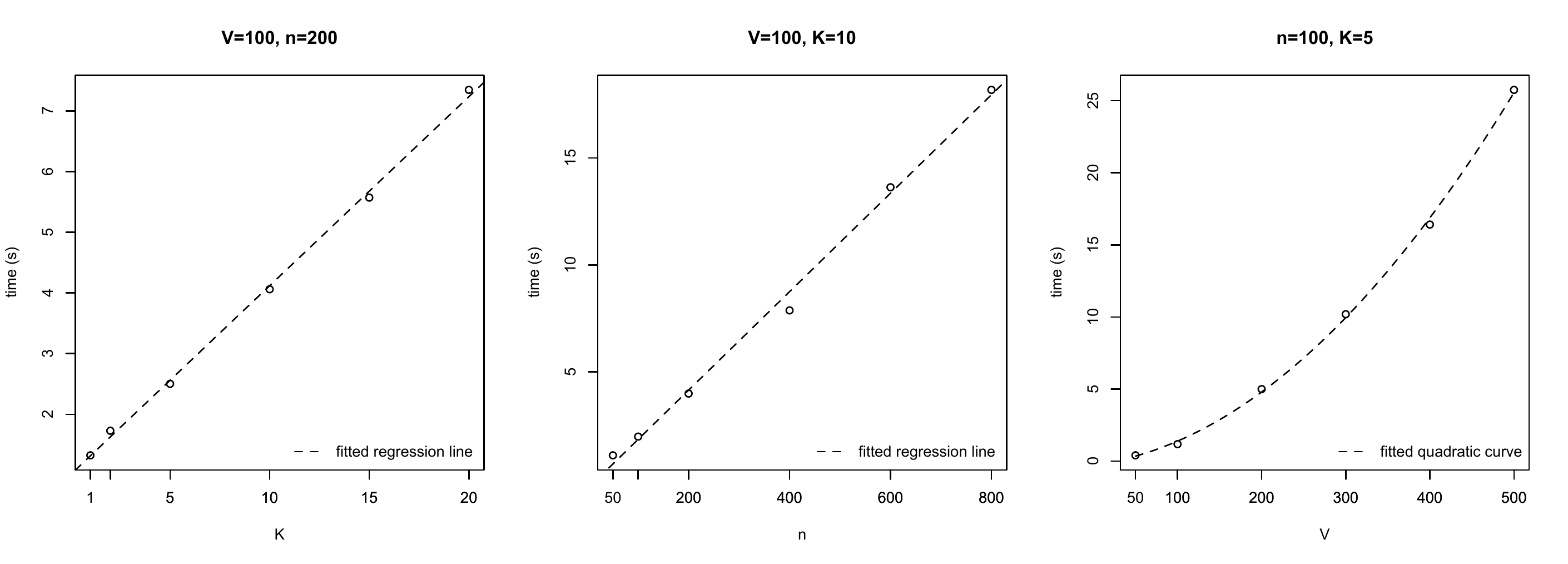}
\caption{Average computation time (in seconds) per iteration of CISE (Algorithm \ref{alg:Joint-inference}) for 30 runs versus latent dimension $K$ (left), number of networks $n$ (middle) and number of nodes $V$ (right). All the numerical experiments are conducted in R (version 3.3.1). \label{fig:run-time}}
\end{figure}

Each iteration of CISE  includes two steps: (1) $L_2$-penalized logistic regression and (2) $n$ partial eigenvalue decompositions of $V \times V$ matrices. We simulated a sequence of Erd\"os-R\'enyi graphs (each edge is present with probability 0.5) for different numbers of nodes and then assess how the execution time increases with the problem size. Figure \ref{fig:run-time} displays the average computation time per iteration of CISE algorithm (in R) as a function of the latent dimension $K$, the number of networks $n$ and the number of nodes $V$. We can see that for large problem size with $n=800$, $V=100$ and $K=10$, each iteration of CISE on average takes less than 20 seconds; with $V=500$, $n=100$ and $K=5$, the average running time is around 25 seconds. The runtime of each CISE iteration in Matlab is similar to that in R though a bit longer for small problem size. 
From Figure \ref{fig:run-time}, it is clearly seen that CISE exhibits a linear order with $K$ and $n$, and a quadratic order with $V$, i.e. $O(V^2nK)$,   which is the same as our theoretical analysis in Algorithm \ref{alg:Joint-inference}.


\begin{figure}[!htb]
\includegraphics[width=\textwidth]{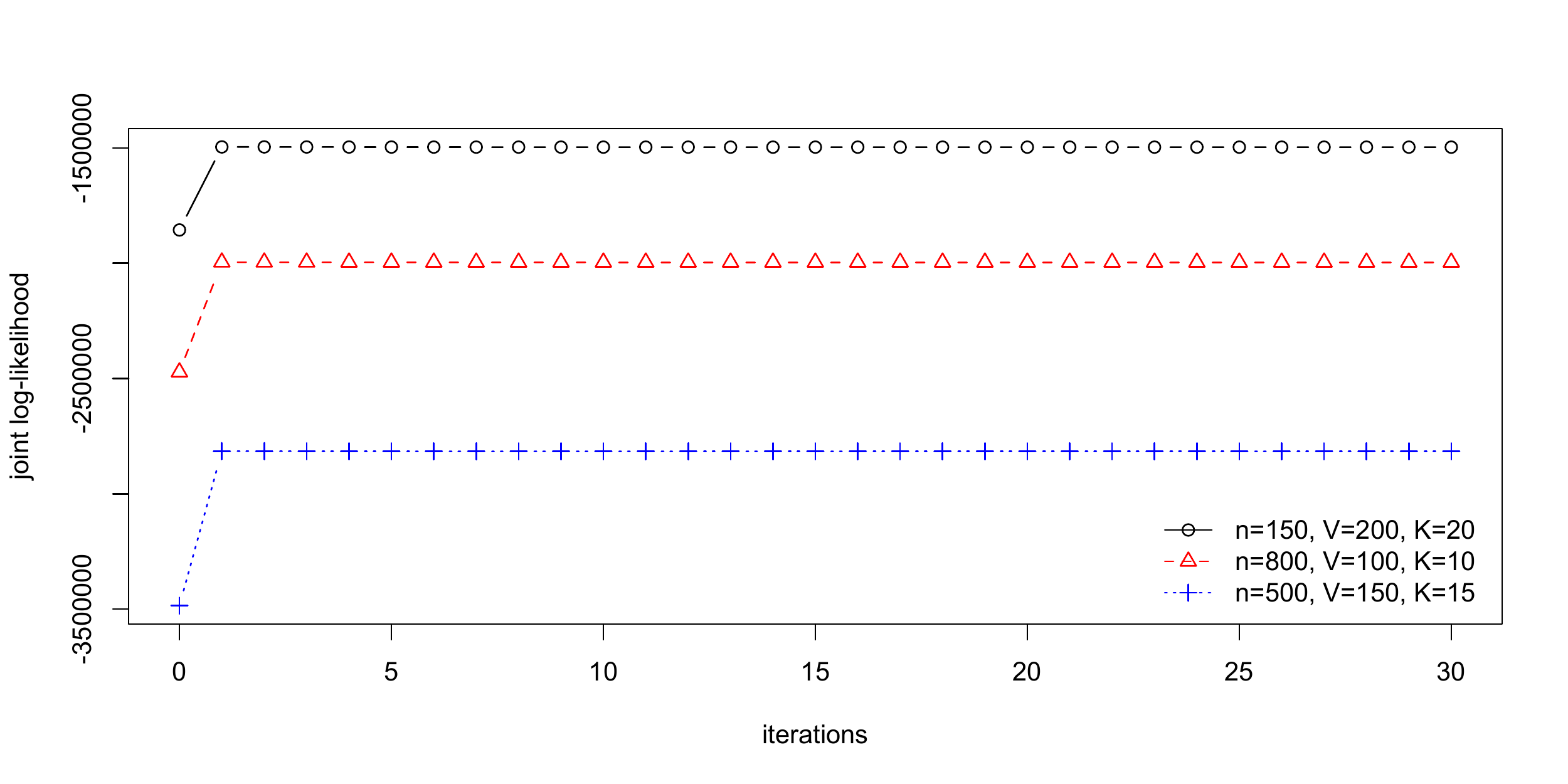}
\caption{CISE algorithm: joint log-likelihood over iterations under different values of $n$, $V$ and $K$. \label{fig:log-likelihood_ER}}
\end{figure}

CISE is a block coordinate descent algorithm, and is guaranteed to converge to a (local) mode. In our experience with simulated and real data, CISE generally converges very fast with a good initialization as specified in Algorithm \ref{alg:Joint-inference}: it usually takes less than 5 steps before the relative change in the joint log-likelihood becomes less than 1\% even for very large problem size. Figure \ref{fig:log-likelihood_ER} shows how the joint log-likelihood \eqref{eq: joint-loglikelihood} evolves over iterations under different problem sizes.  CISE is much more efficient than the Gibbs sampler in \citet{durante2017nonparametric} which conducts Bayesian inference on a related model to M-GRAF but could take hours or days to run for the same problem size. In practice when dealing with real brain network data, we suggest setting $\epsilon$=0.01 in Algorithm \ref{alg:Joint-inference} based on our experiments.

\subsection{Inference on Common and Individual Structure}
\label{simu_infer}

The goal in this section is to assess the performance of our proposed method in terms of inference on the shared versus individual-specific components of variability in replicated networks. To mimic the real brain network data, we first estimate $Z$ and $\{D_i\}$ under $K=3$ from about $800$ $68 \times 68$ structural brain networks extracted from HCP data.  Then the networks are simulated from the M-GRAF model based on the estimated $\hat{Z}$ and $\{\hat{D}_i\}$. 

\begin{figure}[!]
\begin{centering}
\includegraphics[width=0.45\textwidth]{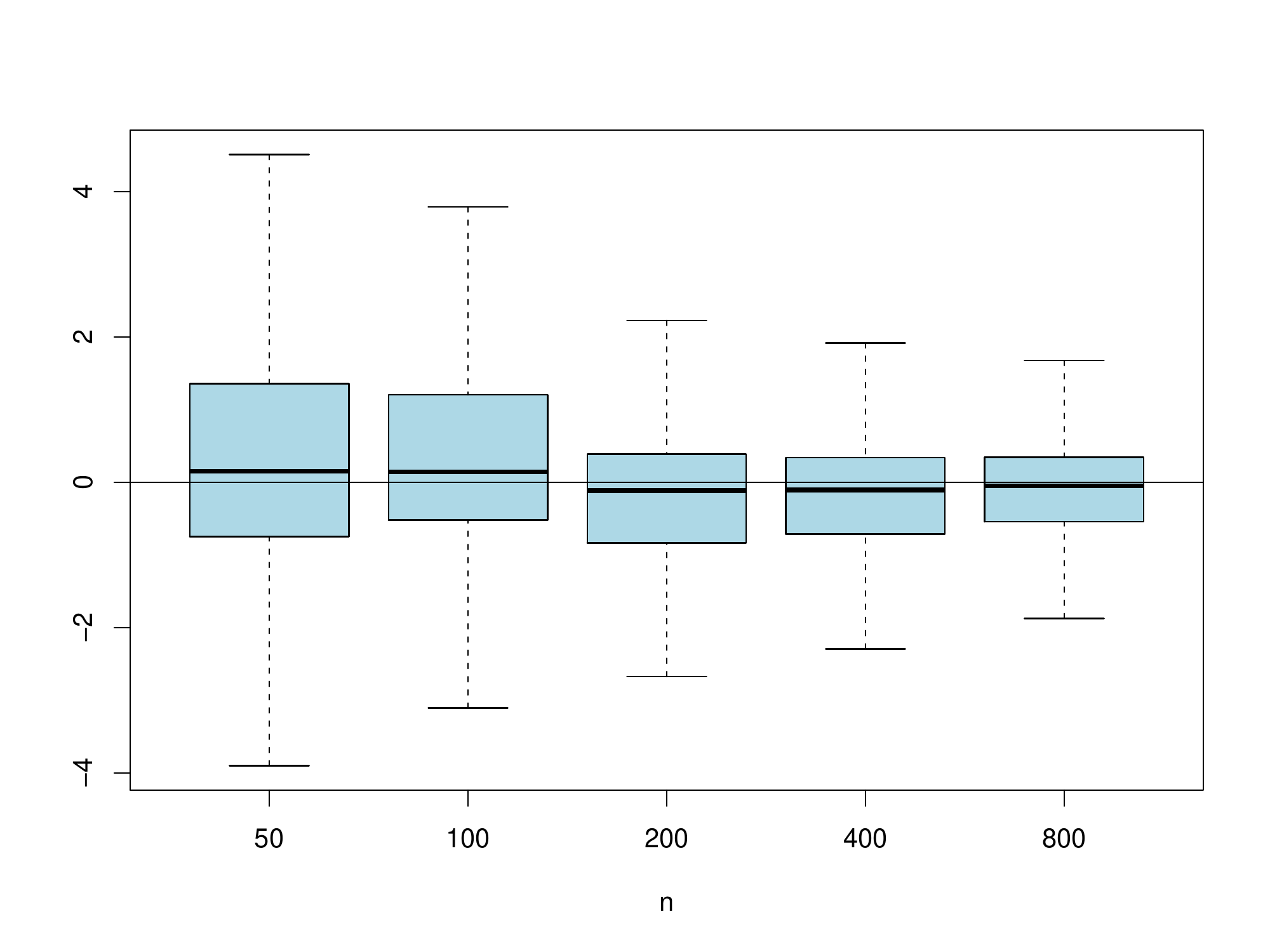}
\includegraphics[width=0.45\textwidth]{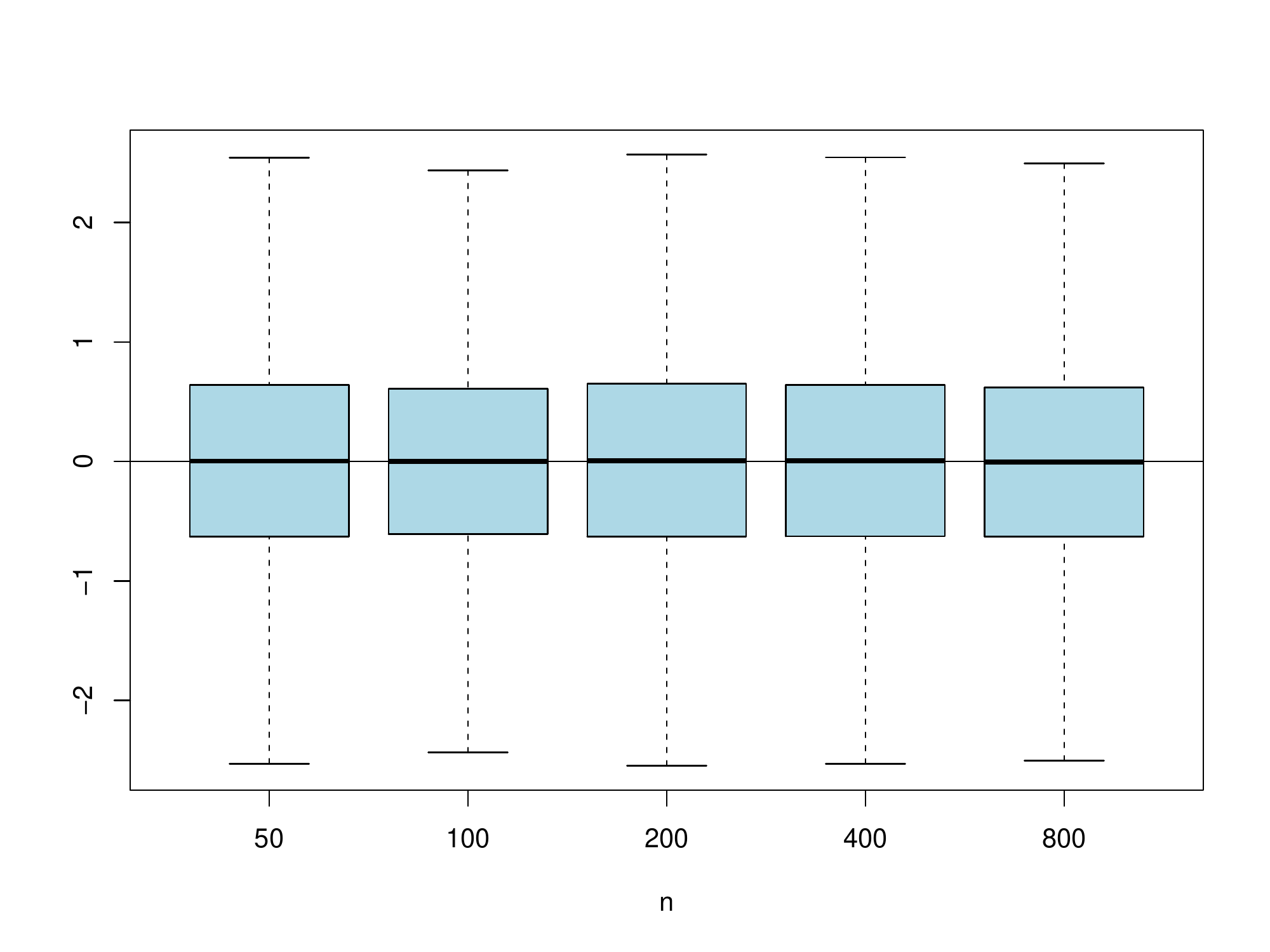}
\par\end{centering}
\caption{Box plots of pooled differences across $50$ simulations between $\mathcal{L}(\hat{Z})$ and $\mathcal{L}(Z_0)$ (left) and $\mathcal{L}(\hat{D}_i) - \mathcal{L}(D_{i0})$ for $20$ randomly selected networks (right) under each number of networks $n$. The networks are generated from M-GRAF model with $K=3$. \label{fig:boxplot_diff}}
\end{figure}

\begin{figure}[!htb]
\includegraphics[width=0.32\textwidth]{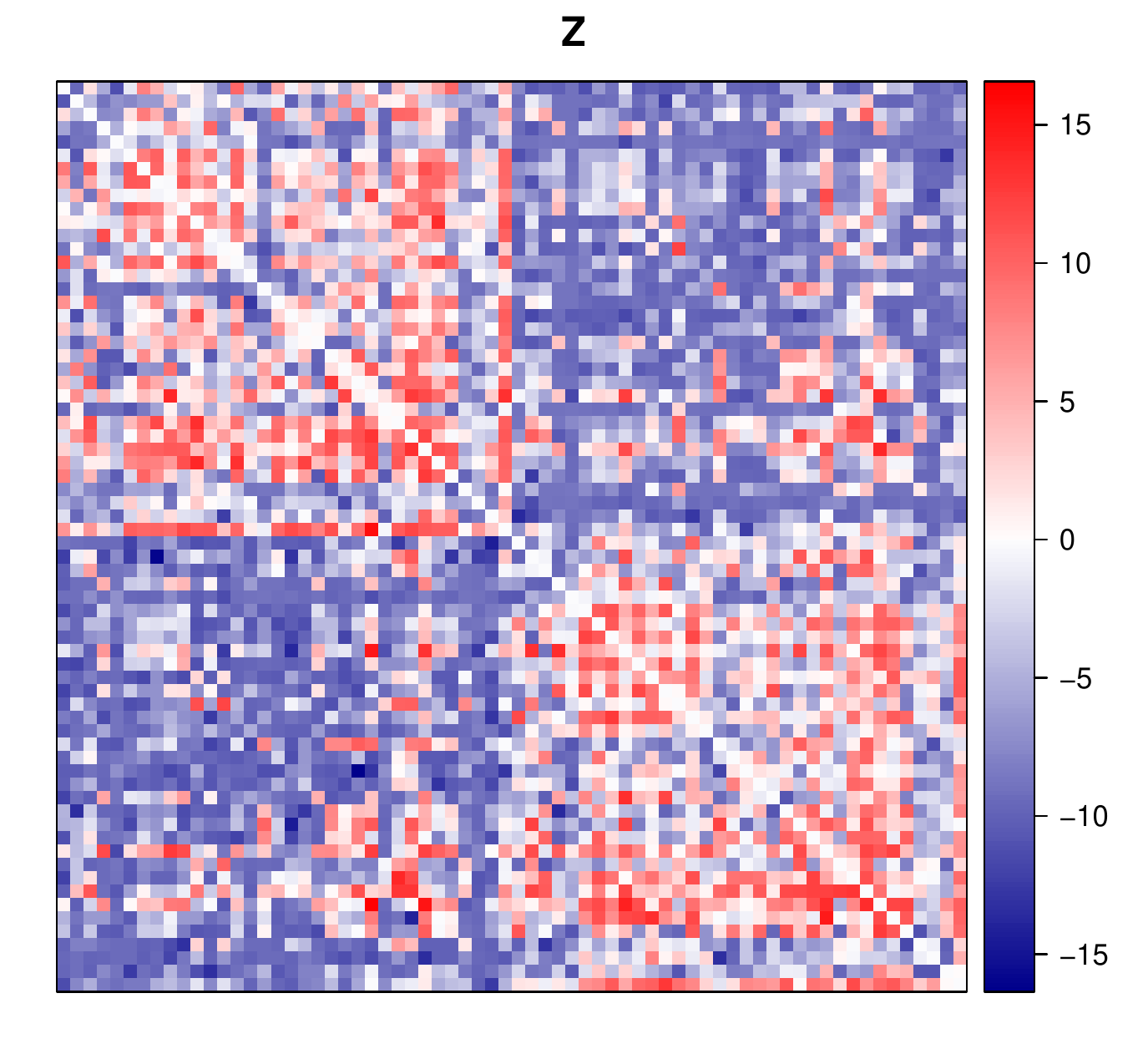}
\includegraphics[width=0.32\textwidth]{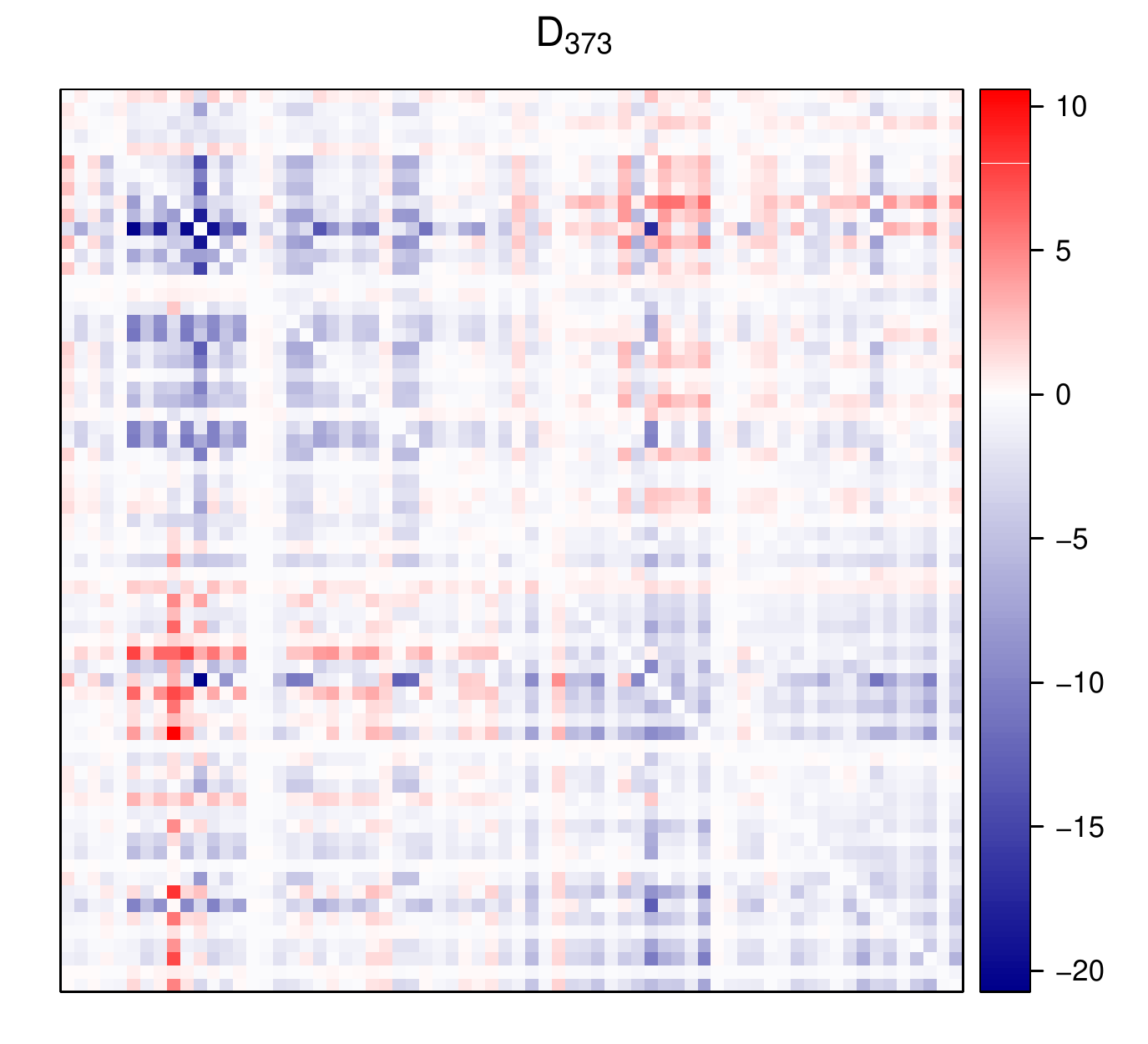}
\includegraphics[width=0.32\textwidth]{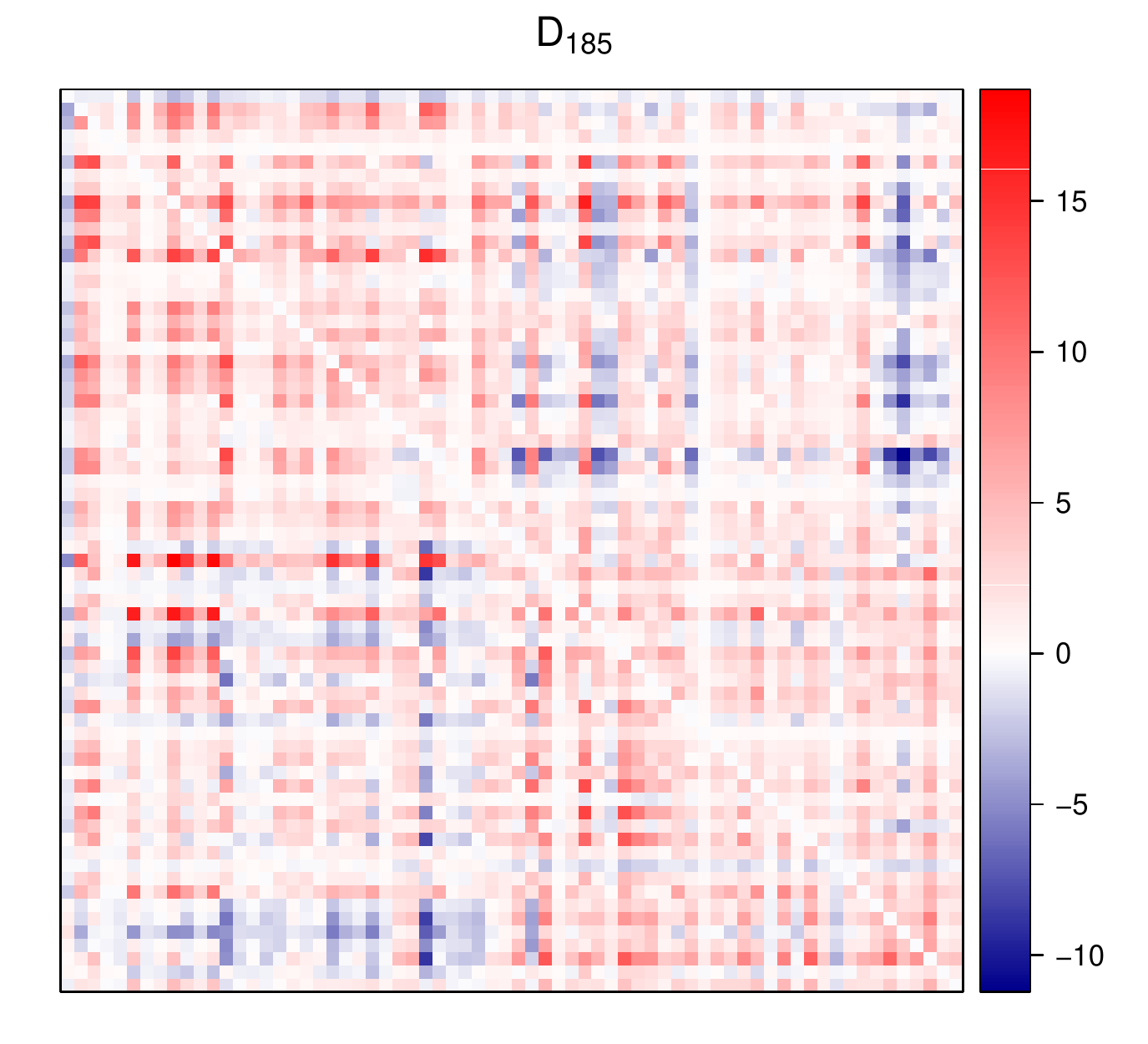}
\caption{Level plots for estimated parameters (lower triangular) versus their true values  (upper triangular) with $n=800$ and $K=3$. Left:  $\hat{Z}$ versus $Z_0$; Middle and Right: $\hat{D}_i$ versus the true $D_{i0}$ for two networks, where the $373$-th network (middle) has the lowest network density and the $185$-th network (right) has the highest network density in the synthetic data. \label{fig:levelplot_ZD}}
\end{figure}

We conduct a sequence of numerical experiments to demonstrate properties of the estimated  parameters in the M-GRAF model as the number of networks grows.  The true values of $Z$ and $\{D_i\}$ are denoted as $Z_0$ and $\{D_{i0}\}$ where each $D_{i0}$ has rank $K=3$. We generate different numbers $n$ ($n=50, 100, 200, 400, 800$) of $68 \times 68$ adjacency matrices from the M-GRAF model based on $Z_0$ and randomly selected $D_{i0}$'s. At each value of $n$, we run CISE algorithm with $K=3$ to obtain the estimated parameters $\hat{Z}$ and $\{\hat{D}_i\}$. Element-wise differences between the lower triangular entries of $\hat{Z}$ and $Z_0$ and the counterpart between $\hat{D}_i$ and $D_{i0}$ for 20 randomly selected networks are recorded. The procedure described above is repeated 50 times where each time we randomly permute 10\% of the entries in $Z_0$. Figure \ref{fig:boxplot_diff} displays boxplots of the pooled differences between estimated parameters and their true values under each $n$ across 50 simulations. Based on the plot, the differences between $\hat{Z}$ and $Z_0$ seem to converge to 0 as $n$ increases. We also notice that the differences  between $\hat{D}_i$ and $D_{i0}$ are centered around 0 and stable across $n$, which is as expected since the number of parameters in $\{D_i\}$ increases with $n$.
Figure \ref{fig:levelplot_ZD} displays the estimated $\hat{Z}$ and $\hat{D}_i$'s versus their corresponding true values from one experiment under $n=800$.

\subsection{Selection of the Dimensionality $K$}

In the above simulation experiments of this section, we assume the dimensionality $K$ is known and simply set $K$ equal to its true value. But in practice, we face a model selection problem. 

\begin{wrapfigure}{r}{0.5\textwidth}
\vspace{0pt}
\begin{centering}
\includegraphics[width=0.5\textwidth]{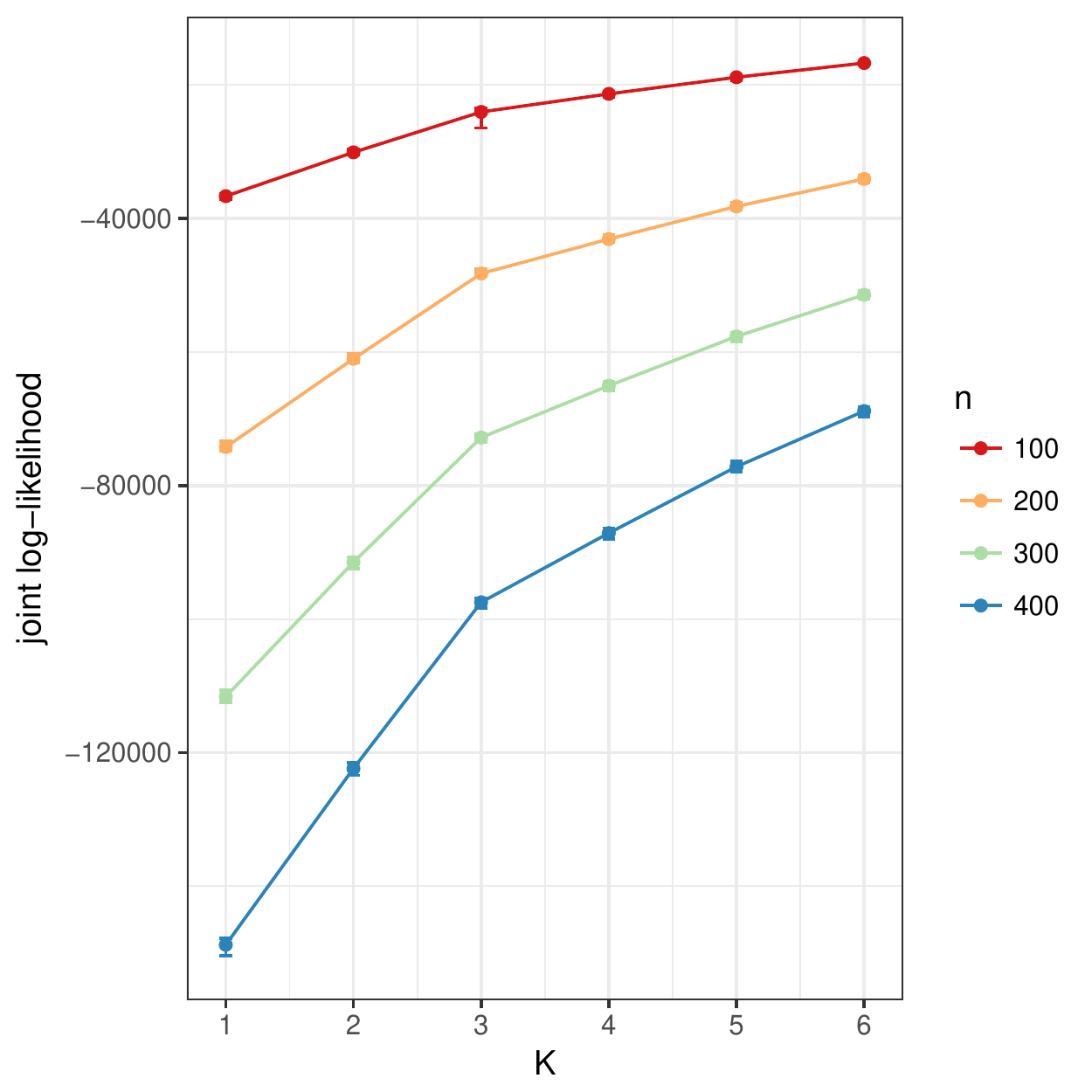}
\end{centering}
\caption{Mean joint log-likelihood at convergence of CISE algorithm with their $95$\% confidence intervals across 50 simulations versus dimension $K$ under different numbers of replicated networks. The networks are simulated from M-GRAF model with $K=3$ as described in Section \ref{simu_infer}. \label{fig:joint LL vs K}}
\vspace{0pt}
\end{wrapfigure}

In the scenario that  we have some extra categorical variable in the dataset and the goal is to do prediction, we can use cross validation to choose $K$ as illustrated in Section \ref{sec:app}. 
Otherwise we recommend the classical ``Elbow'' method to determine $K$, which requires first running CISE algorithm for a sequence of $K$'s and plotting the joint log-likelihood \eqref{eq: joint-loglikelihood} at convergence versus dimension $K$. 
Then the optimal $K$ is determined to be the bend point where the objective function starts to increases slowly as shown in Figure \ref{fig:joint LL vs K}. The plot implies that the bend point is at $K=3$ for different numbers $n$ of networks, which coincides with the true dimension in our data generating process. Based on our study, this approach outperforms AIC or BIC particularly when $n$ is large.

\section{Applications to Structural Brain Networks}
\label{sec:app}

In this section, we apply M-GRAF to two real datasets involving $256$ HCP subjects: (1) HCP scan-rescan dataset and (2)  a subset of HCP $1200$ subjects dataset. Each subject is preprocessed using a state-of-the-art dMRI preprocessing pipeline \citep{Zhang2017HCP} to extract $68 \times 68$ binary structural networks based on the Desikan parcellation atlas \citep{Desikan2006968}. Certainly, even state-of-the-art tractography is subject to measurement errors, but ground truth measurements on actual neurofibers are unavailable given current technology.  Hence, there will be two components of variability in the measured brain networks, one attributed to systematic variability across subjects in their brain connection structure, and one due to measurement errors. Our model can accomodate these two components of variability, with the low-rank assumption on individual deviation not only capturing the main variation of each graph but also serving as a denoising procedure. 

In addition to the network data, we also extract a cognitive trait, measuring the subject's  visuospatial processing ability, to study the relationship between brain connectivity and this cognitive score.

\subsection{Scan-Rescan Brain Network Data}
\label{subsec:scan-rescan}

\begin{figure}
\includegraphics[width=\textwidth]{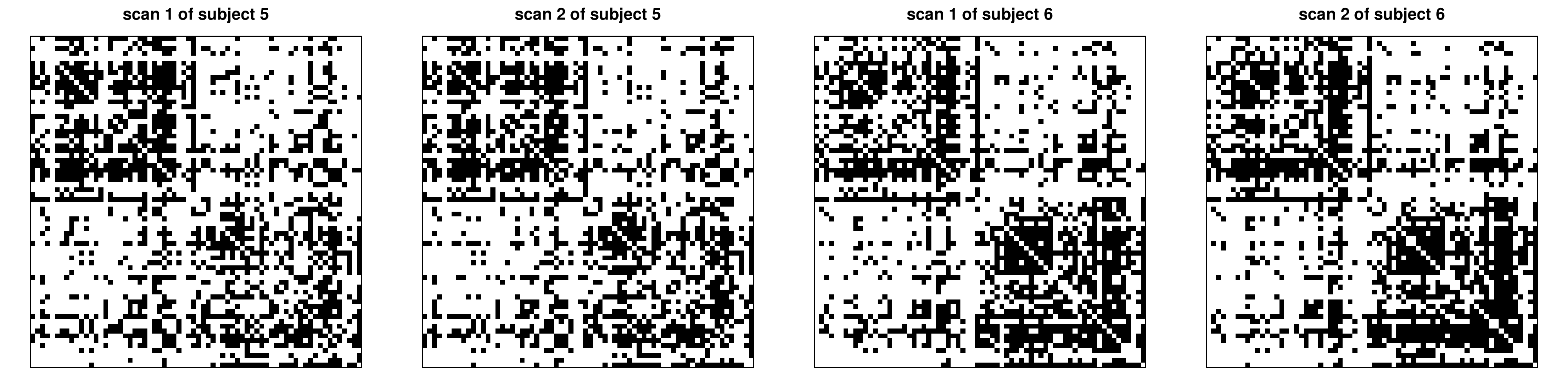} 
\caption{Paired adjacency matrices for two subjects in the HCP scan-rescan data. \label{fig:Paired-adjacency-scan-rescan}}
\end{figure}

In this application, we compare the performance of CISE (Algorithm \ref{alg:Joint-inference}) with several other low-rank approximation methods on the scan-rescan brain network data. The data were collected for 44 healthy subjects under a scan-rescan session, and therefore two $68\times68$ binary adjacency matrices are available for each subject for a total of $n=88$ brain networks. Two examples of the scan-rescan networks extracted for two subjects are shown in Figure \ref{fig:Paired-adjacency-scan-rescan}. It is easy to observe that differences between scan-rescan adjacency matrices for the same subject are much smaller than those between the adjacency matrices for different subjects

These scan-rescan data provide an appealing setting for studying how discriminative the latent structure can be in identification of subjects. The idea is to first learn a low-rank representation for each brain network and then check whether the pairs of networks having the closest low-rank representations correspond to the same subjects.
Specifically, we use the distance measure $d(i,j)$ between scan $i$ and $j$ as introduced in Section \ref{subsec:classification} and then conduct leave-one-out cross validation (LOOCV): for a test subject $i^{\star}$, find $j^{\star}=\underset{j\neq i^{\star}}{\argmin\ }d(i^{\star},j)$ and check if $i^{\star}$ and $j^{\star}$ correspond to the same person. Similarly,  for the model where $D_{i}=Q_{i}\Lambda Q_{i}^{\top}$, the pairwise distance is defined as
\begin{eqnarray*}
d^{2}(i,j) & = & \left\Vert Q_{i}\Lambda Q_{i}^{\top}-Q_{j}\Lambda Q_{j}^{\top}\right\Vert _{F}^2 \\
 & = & 2\tr \left(\Lambda^{2}\right)-2\tr (\Lambda Q_{i}^{\top}Q_{j}\Lambda Q_{j}^{\top}Q_{i}).
\end{eqnarray*}
Another variant of our model $D_{i}=Q\Lambda_{i}Q^{\top}$ does not provide good fit to the data and thus we do not display the results below.

We compare the performance of CISE with some popular matrix and tensor decompositions on multiple network data as below. For a fair comparison, we apply these low-rank approximation methods to the demeaned adjacency matrices $\{A_{i}-\bar{A}:i=1,\dots,n\}$ where $\bar{A}=\sum A_{i}/n$, so as to better capture the deviation of each network from their common structure.
\begin{itemize}
\item Separate factorization. We apply the spectral embedding method \citep{sussman2012consistent} separately to each network in the dataset where each probability matrix $\Pi_{i}$ is estimated by the sum of $\bar{A}$ and a low rank approximation to $(A_{i}-\bar{A})$ via SVD.

\item CP decomposition. Let $\boldsymbol{{\cal A}}_d$ denote the $V\times V\times n$ tensor of demeaned adjacency matrices. The CP decomposition seeks to model $\boldsymbol{{\cal A}}_d$ as a sum of rank-one tensors: $\boldsymbol{{\cal A}}_d \approx\sum_{k=1}^{K}d_{k}\boldsymbol{u}_{k}\circ\boldsymbol{v}_{k}\circ\boldsymbol{w}_{k}$, where $\boldsymbol{u}_{k}\in\mathbb{R}^{V}$, $\boldsymbol{v}_{k}\in\mathbb{R}^{V}$, $\boldsymbol{w}_{k}\in\mathbb{R}^{n}$, $d_{k}\geq0$ and $\circ$ denotes the outer product \citep{kolda2009tensor}. Unlike the singular value decomposition (SVD) for a matrix, CP decomposition does not uniquely decompose the data \citep{kolda2009tensor}, which may complicate the analysis. Similar to \citet{sussman2012consistent}, each probability matrix $\Pi_{i}$ is estimated by $\hat{\Pi}_{i}=\bar{A} + \sum_{k=1}^{K}d_{k}w_{ki}\boldsymbol{u}_{k}\boldsymbol{v}_{k}^{\top}$, where $w_{ki}$ is the $i$th entry of $\boldsymbol{w}_{k}$.

\item Tucker decomposition. Tucker decomposition seeks to model $\boldsymbol{{\cal A}}_d$ as $\boldsymbol{{\cal A}}_d \approx\boldsymbol{{\cal D}}\times_{1}U_{1}\times_{2}U_{2}\times_{3}W$ where $\boldsymbol{{\cal D}}$ is a $K_{1}\times K_{2}\times K_{3}$ core tensor and the factors $U_{1}\in\mathbb{R}^{V\times K_{1}}$, $U_{2}\in\mathbb{R}^{V\times K_{2}}$ and $W\in\mathbb{R}^{n\times K_{3}}$ are orthonormal matrices \citep{kolda2009tensor}. We set $K_{1}=K_{2}=K$ and $K_{3}=n$ in this case, and again we consider each matrix along the 3rd dimension of the low rank tensor plus $\bar{A}$ as the estimated probability matrix $\hat{\Pi}_{i}$.
\end{itemize}

We use R package \texttt{rTensor} to compute the components in tensor decompositions. The distance measure in these methods is defined as $d(i,j)\coloneqq\Vert \hat{\Pi}_{i}-\hat{\Pi}_{j}\Vert _{F}$. We report the LOOCV accuracy of subject identification on the scan-rescan data in Table \ref{tab:scan-rescan_pred_accu}. The results show that the accuracy from the variant of our model $D_{i}=Q_{i}\Lambda Q_{i}^{\top}$ is always the highest under the same rank $K$ and reaches 1 at $K=8$. Although separate factorization has the same accuracy as our model at rank 2 and 5, its accuracy increases more slowly with $K$. The two tensor decomposition methods have poor classification performance here, implying that their low rank approximations are not discriminative enough in this scenario.
\begin{table}[!htb]
\centering
\caption{LOOCV identification accuracy on scan-rescan data for different methods. \label{tab:scan-rescan_pred_accu}}
\begin{tabular}{c >{\centering}m{2cm} >{\centering}m{1.9cm} >{\centering}m{1.8cm}  >{\centering}m{1.8cm}  >{\centering}m{1.8cm} }
\toprule 
 & M-GRAF1 $D_{i}=Q_{i}\Lambda_{i}Q_{i}^{\top}$ & M-GRAF2 $D_{i}=Q_{i}\Lambda Q_{i}^{\top}$ & Separate factorization & CP\\ decomposition & Tucker\\ decomposition\tabularnewline
\midrule
$K=2$ & 0.705 & \textbf{0.761} & \textbf{0.761} & 0.114 & 0.136\tabularnewline
$K=5$ & 0.886 & \textbf{0.932} & \textbf{0.932} & 0.477 & 0.670\tabularnewline
$K=7$ & 0.966 & \textbf{0.989} & 0.943 & 0.591 & 0.761\tabularnewline
$K=8$ & 0.977 & \textbf{1.000} & 0.966 & 0.625 & 0.841\tabularnewline
\bottomrule
\end{tabular}
\end{table}

After obtaining a discriminative latent structure, we want to further check how well edges in the networks can be predicted. We compute the area under the ROC curve (AUC) in predicting ${\cal L}(A_{i})$ with estimated probability matrix $\hat{\Pi}_{i}$ and the residual sum of squares (RSS), i.e. the $L_{2}$-norm of the difference between ${\cal L}(A_{i})$ and ${\cal L}(\hat{\Pi}_{i})$. The mean and standard deviation of AUC and RSS across all the subjects are reported in Table \ref{tab:AUC and MSE}, which shows that CISE has higher AUC and lower RSS than other methods with the same rank $K$. 
The results from the two variants of our model are quite similar, though allowing $\Lambda_{i}$ to vary across individuals performs slightly better due to more flexibility. 
\begin{table}[!hbt]
\centering
\caption{Mean and standard deviation of AUC and RSS across subjects under different $K$'s.\label{tab:AUC and MSE}}
\begin{tabular}{c>{\centering}m{2.5cm}>{\centering}m{2cm}>{\centering}m{2cm}>{\centering}m{2cm}>{\centering}m{2cm}}
\toprule 
 & M-GRAF1 $D_{i}=Q_{i}\Lambda_{i}Q_{i}^{\top}$ & M-GRAF2 $D_{i}=Q_{i}\Lambda Q_{i}^{\top}$ & Separate factorization & CP\\ decomposition & Tucker\\ decomposition\tabularnewline
\midrule 
\multicolumn{6}{c}{AUC}\tabularnewline
\midrule
$K=2$ & \textbf{0.9880$\pm$0.0024} & 0.9877$\pm$0.0024 & 0.9846$\pm$0.0031 & 0.9758$\pm$0.0043 & 0.9758$\pm$0.0043\tabularnewline
$K=5$ & \textbf{0.9948$\pm$0.0014} & 0.9945$\pm$0.0014 & 0.9928$\pm$0.0017 & 0.9768$\pm$0.0042 & 0.9777$\pm$0.0040\tabularnewline
$K=7$ & \textbf{0.9969$\pm$0.0009} & 0.9968$\pm$0.0009 & 0.9959$\pm$0.0011 & 0.9774$\pm$0.0037 & 0.9791$\pm$0.0037\tabularnewline
$K=8$ & \textbf{0.9976$\pm$0.0008} & 0.9974$\pm$0.0007 & 0.9970$\pm$0.0008 & 0.9779$\pm$0.0036 & 0.9800$\pm$0.0037\tabularnewline
\midrule
\multicolumn{6}{c}{$\left\Vert {\cal L}(A_{i})-{\cal L}(\hat{\Pi}_{i})\right\Vert _{2}$}\tabularnewline
\midrule
$K=2$ & \textbf{9.63$\pm$0.51} & 9.68$\pm$0.52 & 10.75$\pm$0.46 & 11.62$\pm$0.51 & 11.61$\pm$0.51\tabularnewline
$K=5$ & \textbf{7.65$\pm$0.53} & 7.75$\pm$0.53 & 9.63$\pm$0.40 & 11.50$\pm$0.50 & 11.41$\pm$0.49\tabularnewline
$K=7$ & \textbf{6.64$\pm$0.48} & 6.72$\pm$0.52 & 8.99$\pm$0.37 & 11.42$\pm$0.47 & 11.27$\pm$0.48\tabularnewline
$K=8$ & \textbf{6.18$\pm$0.50} & 6.34$\pm$0.50 & 8.69$\pm$0.36 & 11.38$\pm$0.47 & 11.19$\pm$0.48\tabularnewline
\bottomrule
\end{tabular}
\end{table}

\begin{figure}[!htb]
\includegraphics[width=\textwidth]{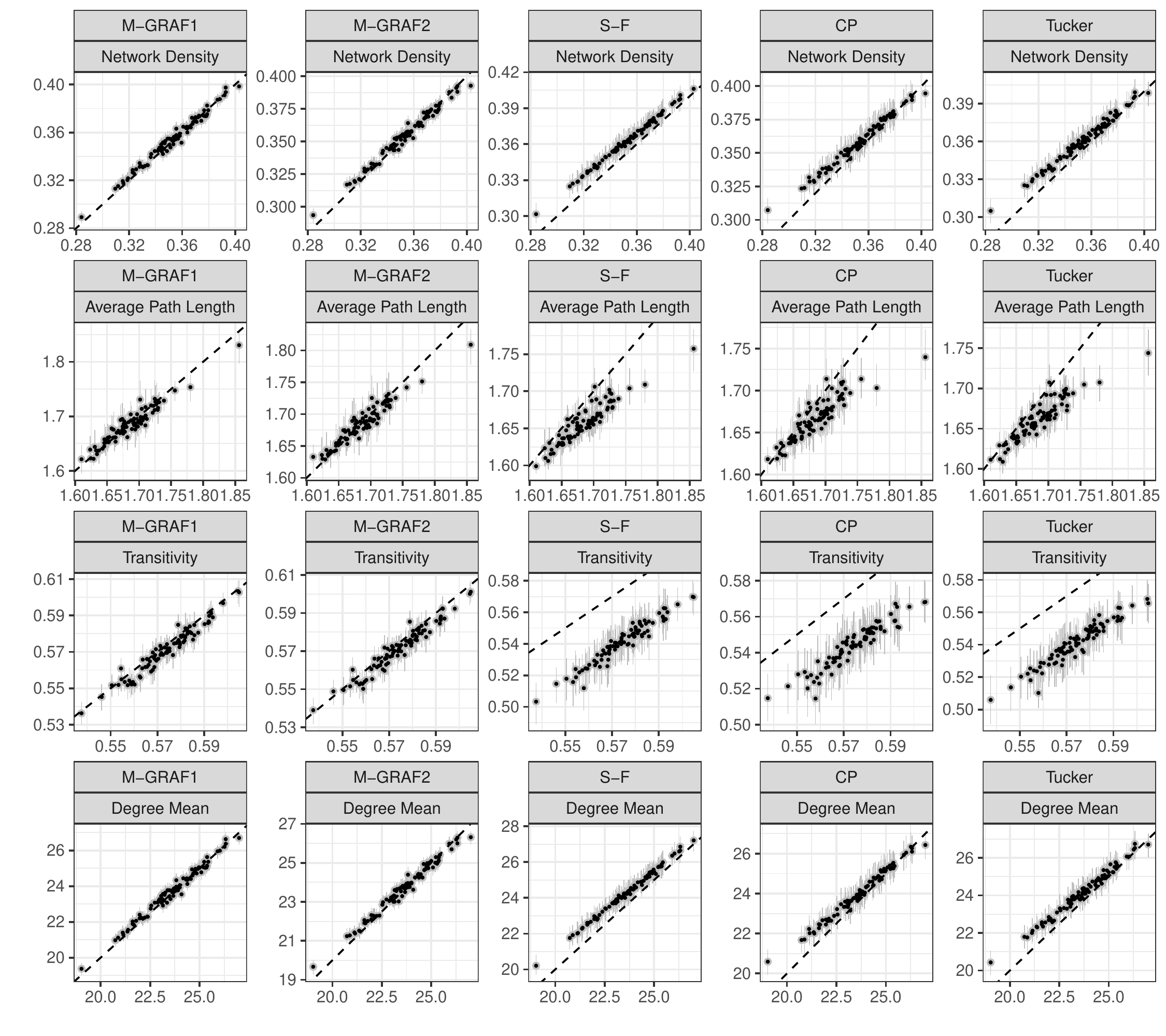}
\caption{ {\small Goodness-of-fit assessment for selected network topological features under different methods. The methods from left to right: M-GRAF1 is M-GRAF with $D_{i}=Q_{i}\Lambda_{i}Q_{i}^{\top}$ under $K=8$; M-GRAF$2$ is M-GRAF with $D_{i}=Q_{i}\Lambda Q_{i}^{\top}$ under $K=8$; S-F is separate factorization with $K=17$; CP decomposition with $K=100$ and Tucker decomposition with $K=36$. The topological features from top to bottom are network density, average shortest path length, transitivity and degree mean. Each dot of the scatterplot corresponds to a subject, where $x$-coordinate denotes her observed topological feature, $y$-coordinate denotes the corresponding predictive mean and the grey segment denotes the 95\% predictive confidence interval. The dashed line in each scatterplot denotes the $y=x$ line. }\label{fig:topo_features}}
\end{figure}

We assess goodness-of-fit by comparing some key topological features of networks observed in the data to those estimated from different methods. The selected topological measures include network density, average shortest path length, transitivity and mean of node degrees (degree mean) \citep{newman2010networks}. Specifically, we first obtain the predictive distributions of these topological measures for each subject by simulating 100 networks from the estimated $\hat{\Pi}_{i}$ under different models, and then compare the predictive means to the empirical topological features via scatterplots along with 95\% confidence intervals as shown in Figure \ref{fig:topo_features}. Each dot in these scatterplots corresponds to a subject with $x$-coordinate being her empirical topological measure and $y$-coordinate the predictive mean. The closer the points are to the dashed diagonal line, the better fit of the model.  
For a fair comparison,  we choose $K=17$ for Separate factorization and $K=36$ for Tucker decomposition in Figure \ref{fig:topo_features} since these choices of $K$ provide an accuracy of 1 for the two methods in the scan-rescan classification task. We set $K=100$ for CP decomposition, which provides an accuracy of around 0.989. 
Figure \ref{fig:topo_features} shows that the two variants of our model provide much better characterization of network topological features than the other methods. In addition, the variant $D_{i}=Q_{i}\Lambda Q_{i}^{\top}$ provides almost indistinguishable predictive results from those under the M-GRAF model with $D_{i}=Q_{i}\Lambda_{i}Q_{i}^{\top}$. 
Therefore, restricting $\Lambda_{i}$ to be the same across subjects seems to be a reasonable assumption for brain network data.

\subsection{Brain Networks and Cognitive Traits}
\label{sec:vp}

The HCP collects measurements on a range of motor, sensory, cognitive and emotional processes for each participant, with an overarching goal being improved understanding of the relationship between brain connectivity and human traits \citep{barch2013function}. For sake of clarity and brevity, we focus here on studying relationships between brain structural connectivity and one particular trait -- visuospatial processing. 

Visuospatial processing is commonly assessed using the Variable Short Penn Line Orientation Test (VSPLOT), where two line segments are presented on the screen and participants are asked to rotate a movable line so that it is parallel to the fixed line; for more details, we refer the readers to \citet{moore2015development}. The latest released HCP data contain VSPLOT scores of about 1200 healthy adults. We preselected subjects having high (top 10\%) and low (bottom 10\%) VSPLOT scores with 106 subjects in each group. Hence the resulting dataset contains an indicator of high/low visuospatial processing score $l_i \in \{0,1\}$ and an adjacency matrix $A_i$ representing the structural connectivity among 68 brain regions for $212$ individuals.

\begin{figure}[!htb]
\includegraphics[width=0.47\textwidth]{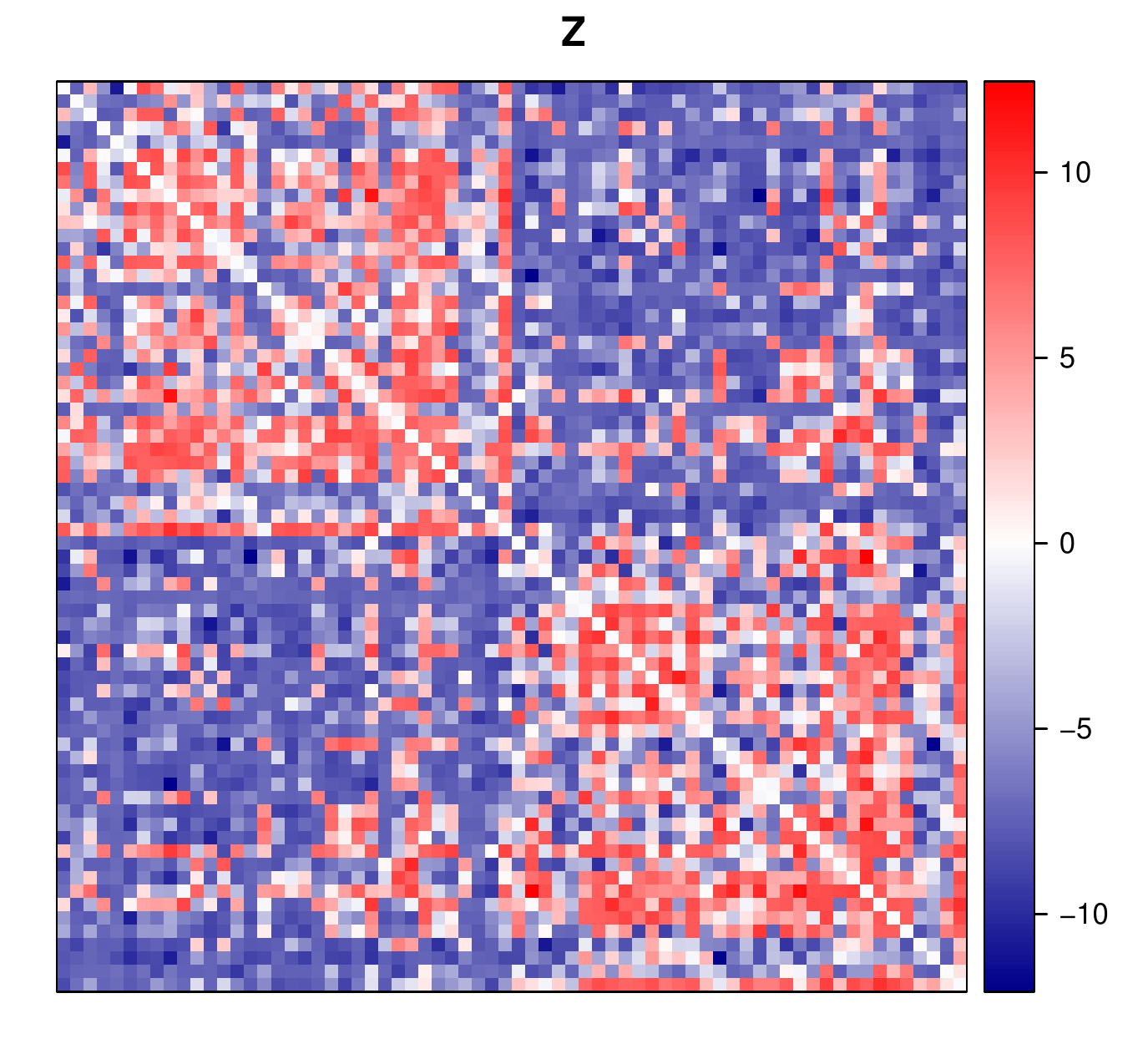}\includegraphics[width=0.53\textwidth]{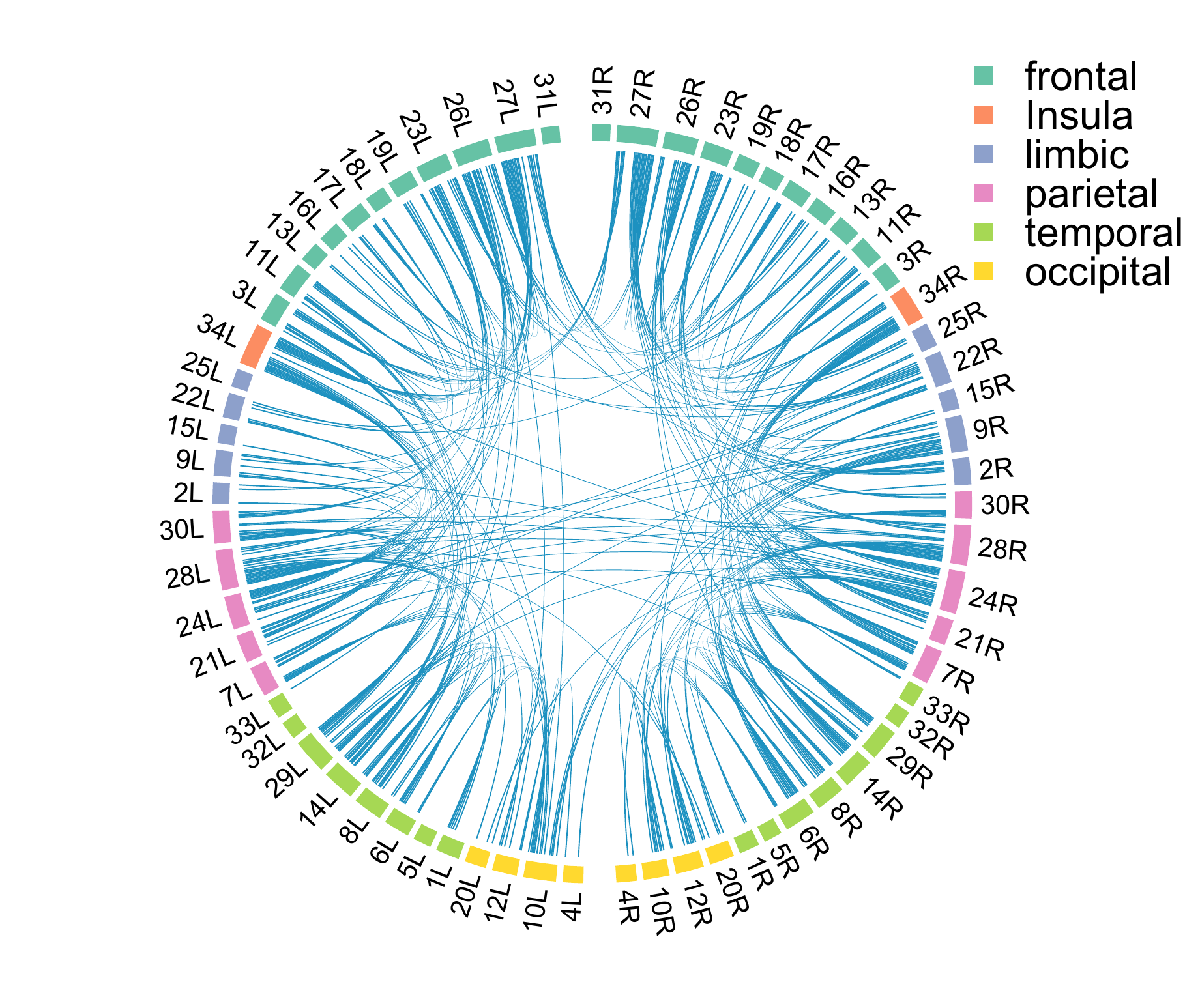}
\caption{Heatmap of the estimated $\hat{Z}$ under $K=5$ (left) and a chord diagram of the connections $uv$'s ($277$ in total) with $\pi(\hat{Z}_{uv}) > 0.999$ (right), where $\pi(\cdot)$ is the logistic function. \label{fig:Z-visuo}}
\end{figure}

We followed the same goodness-of-fit assessment procedure as described in Section \ref{subsec:scan-rescan} and observed very similar performance between the models $D_{i}=Q_{i}\Lambda Q_{i}^{\top}$ and $D_{i}=Q_{i}\Lambda_{i}Q_{i}^{\top}$. Therefore, we choose the variant $D_{i}=Q_{i}\Lambda Q_{i}^{\top}$ to further reduce the number of parameters. We use the distance described in Section \ref{subsec:classification} to classify subjects with high and low visuospatial processing score using their estimated low-rank components $\{Q_{i}\}$ and $\Lambda$. The prediction accuracy is measured by repeating 10-fold cross validation (CV) 30 times. We report the mean and standard deviation of the CV accuracies under different choices of $K$ in Table \ref{tab:Accuracy}. It seems that $K=5$ is enough to provide a good prediction accuracy of 0.643 on average, implying that individual-specific components of brain connectivity are related to visuospatial processing. The estimated common structure $\hat{Z}$ of brain connectivity underlying all subjects is displayed via a heatmap in Figure \ref{fig:Z-visuo}. The chord diagram in Figure \ref{fig:Z-visuo} shows the selected $277$ edges with $\pi(\hat{Z}_{uv}) > 0.999$, where $\pi(\cdot)$ is the logistic function. Hence we expect these connections to be present with probability almost $1$ for an average brain of $212$ HCP subjects.
\vspace{-10pt}
\begin{wraptable}{r}{0.5\textwidth}
\caption{Mean and standard deviation of prediction accuracies in repeated 10-fold cross validation.}
\label{tab:Accuracy}
\centering
\vspace{0pt}
\begin{tabular}{cc}
\toprule 
$K$ & Accuracy\tabularnewline
\midrule
1 & 0.561$\pm$0.101\tabularnewline
2 & 0.621$\pm$0.107\tabularnewline
3 & 0.622$\pm$0.104\tabularnewline
4 & 0.623$\pm$0.105\tabularnewline
5 & \textbf{0.643$\pm$0.102} \tabularnewline
6 & 0.641$\pm$0.104\tabularnewline
7 & 0.629$\pm$0.105\tabularnewline
\bottomrule
\end{tabular}
\end{wraptable}

\begin{figure}[!htb]
\includegraphics[width=0.49\textwidth]{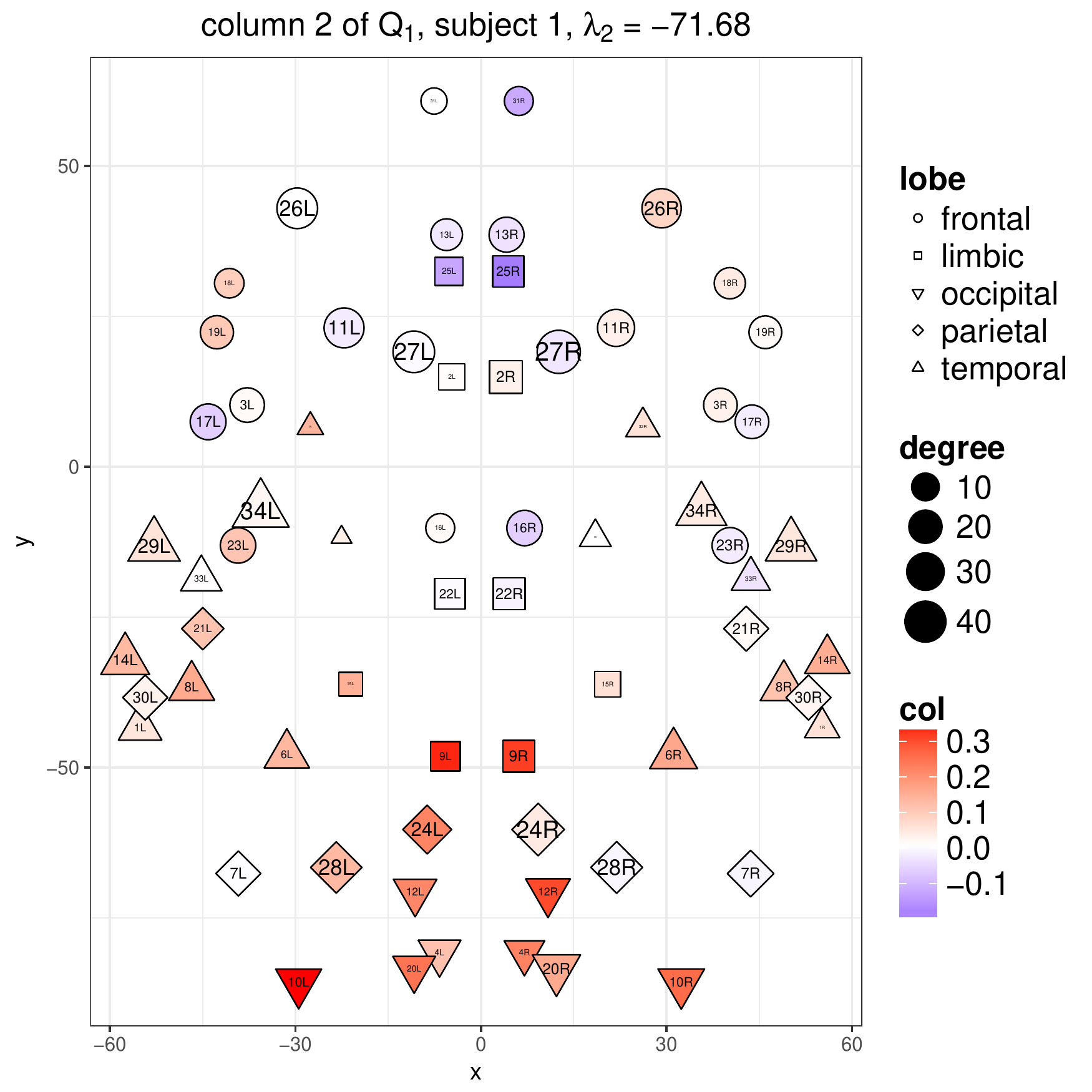}\includegraphics[width=0.49\textwidth]{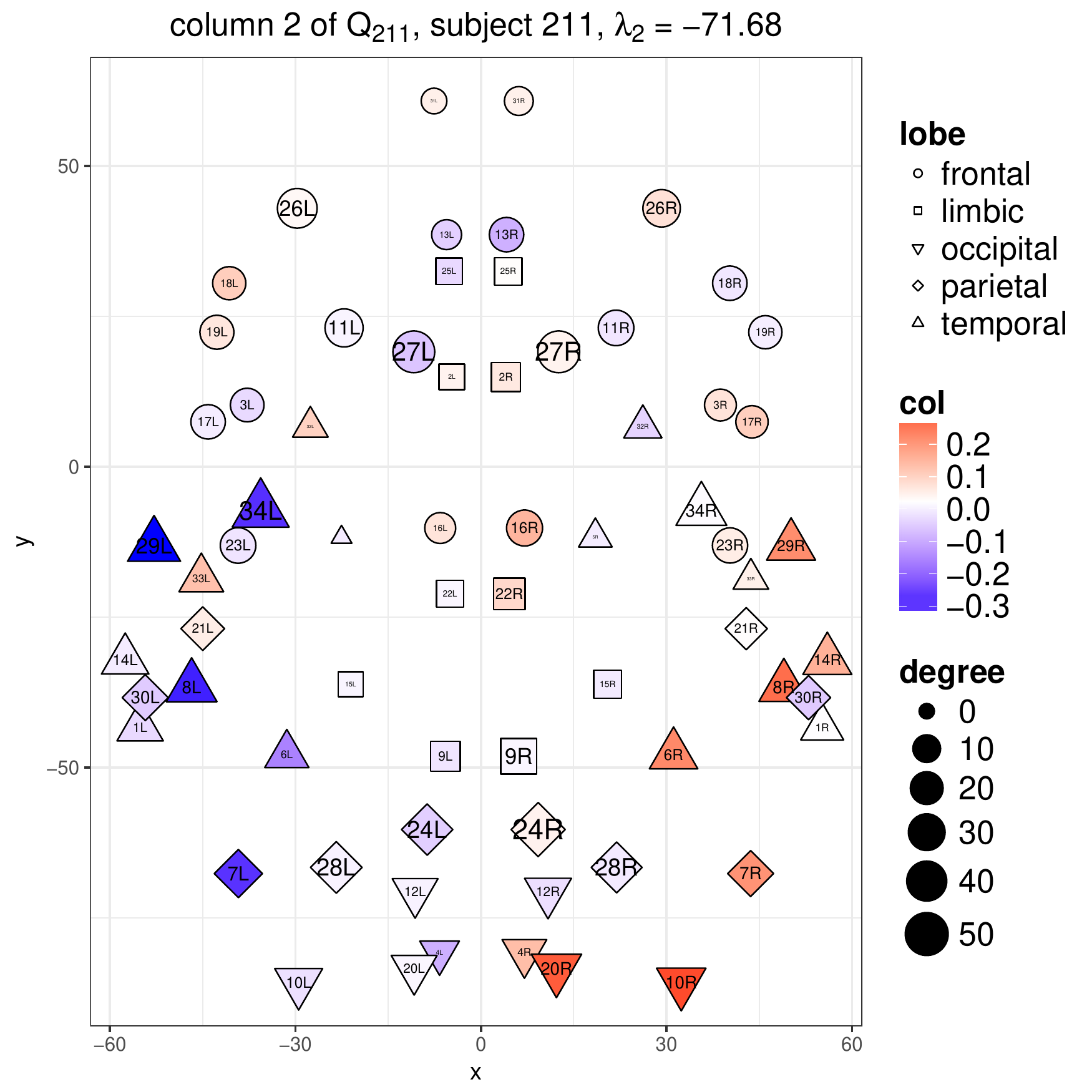}
\caption{Heatmap of the second column of $Q_{i}$ for subject $i=1$ (left) and $i=211$ (right) under $K=2$. Their adjacency matrices are presented in Figure \ref{fig:adj-visuo}. \label{fig:node_heatmap}}
\end{figure}

Table \ref{tab:Accuracy} also shows that $K=2$ leads to a jump in performance relative to $K=1$.  Since $\hat{\lambda} = 83.6$ under $K=1$ and $(\hat{\lambda}_1, \hat{\lambda}_2)  = (77.6, -71.7)$ under $K=2$, we display the second column of $Q_i$ corresponding to $\lambda_2$ via a heatmap across the 68 brain regions for two subjects in Figure \ref{fig:node_heatmap} (their adjacency matrices are shown in Figure \ref{fig:adj-visuo}). According to \citet{moore2015development}, visuospatial processing is linked to posterior cortical function and thus we focus on the regions in the occipital lobe, which is located in the posterior portion of the human cerebral cortex and is the visual processing center of the brain containing most of the anatomical regions of the visual cortex \citep{zeki1991direct}. Subject 1 in the left plot of Figure \ref{fig:node_heatmap} has the lowest score in VSPLOT and we can see that her brain regions located in the occipital lobe (bottom of the plot) all have similar positive coordinates. Since $\hat{\lambda}_2<0$, this indicates that Subject 1 tends to have few connections within the occipital lobe. Subject 211 in the right plot of Figure \ref{fig:node_heatmap} has the highest score in VSPLOT and the coordinates of her brain regions in occipital lobe are not similar, indicating more connections within this lobe. 

\begin{figure}[!htb]
\includegraphics[scale=0.45]{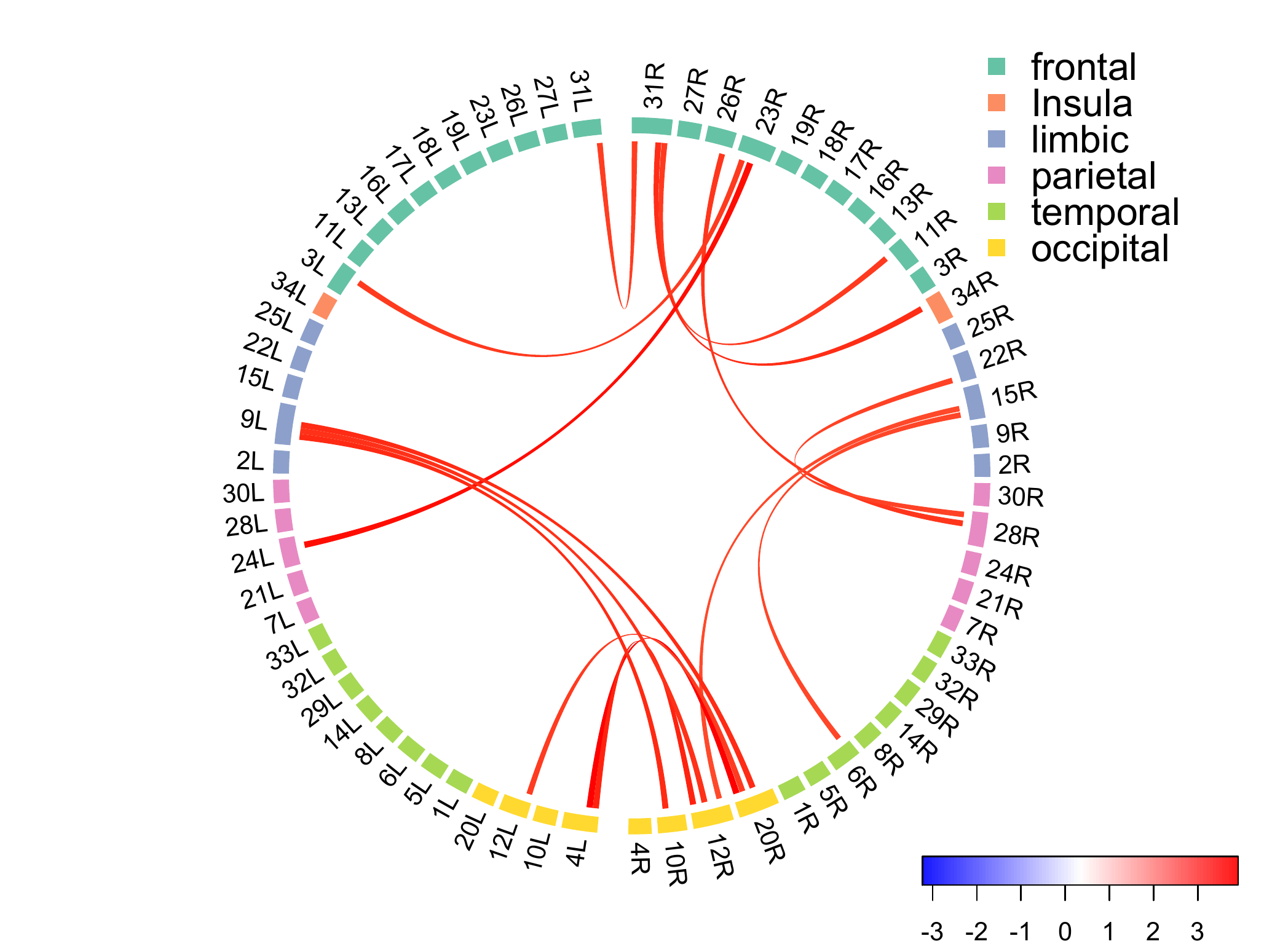}
\caption{Selected subnetwork that might be related to the visuospatial processing: significant connections ($15$ in total) in the $t$-test of $D_{i[uv]}$'s between high and low visuospatial processing group for each edge $uv$ under FDR $\leq 0.15$. The color of the chords represents the corresponding $t$ statistic, which goes from blue to red as $t$ statistic goes from $-3.20$ (minimum among all connections) to $3.88$ (maximum).
\label{fig:Chord-diagram}}
\end{figure}

To identify a subnetwork that might relate to visuospatial processing, we test for differences in the log odds of each connection between the two groups. Specifically, for each connection $uv$ in the brain network, we applied a $t$-test on the $D_{i[uv]}$'s in high and low visuospatial functioning groups under $K=5$. We adjusted for multiple comparisons by rejecting all local nulls having a $p$-value below the \citet{benjamini1995controlling} threshold to maintain a false discovery rate FDR$\le 0.15$. The significant connections are displayed via a chord diagram in Figure \ref{fig:Chord-diagram}. 
Figure \ref{fig:Chord-diagram} shows that many connections in the selected subnetwork relate to regions in the occipital lobe, especially the right occipital lobe. This seems consistent with neuroimaging and lesion studies which provide evidence of dysfunction in right posterior regions of the brain for deficits in visuospatial processing \citep{moore2015development}. In particular in the occipital lobe, Region $12R$ (right lingual) and $20R$ (right peri calcarine) in Figure \ref{fig:Chord-diagram} seem to be the most affected regions related to visuospatial processing since they have more connections with differences between the two groups. This agrees with the findings that damage to the lingual gyrus leads to a form of topographic disorientation \citep{kravitz2011new} and abnormalities in calcarine sulcus, which is a key node of the ventral visual pathway, are related to impaired visual information processing \citep{wu2015cortical}.

\section{Conclusion}
\label{sec:conc}

In this paper, we develop a framework for studying common and individual structure of multiple binary undirected networks with similar patterns. Two variants of the model have been proposed to account for different degrees of heterogeneity in the data so as to avoid overfitting. We have developed an efficient algorithm - CISE - for estimating the model based on spectral decomposition. Simulation studies have illustrated the fast computation of CISE algorithm on large datasets and good properties in inference. We also demonstrated how accounting for common structure can lead to a much lower dimensional individual latent structure, which is highly discriminative in scan-rescan identification. Our approach also provides better prediction and goodness-of-fit (in terms of topological properties) to brain network data than some popular dimension-reduction methods.

Although CISE algorithm has good performance when the latent dimension $K$ is small, it can get trapped at some local modes when $K$ is large due to  high dimensionality of the parameter space. A multi-resolution approach might be a solution to this issue, where we apply a coarse to fine factorization of Z and the estimates of the parent entries in the previous layer provide prior information for the daughter entries in the next layer. This technique may prevent some parameters from getting trapped in local modes leading to a better optima.

\appendix

\section{Proofs of Propositions} 
\label{app}

This appendix contains proofs of Proposition \ref{prop:trace} and Proposition \ref{prop:prop2} in Section \ref{subsec:MLE_Q} as well as an algorithm for inference of $Q$ in the variant $D_{i}=Q\Lambda_{i}Q^{\top}$.

\subsection{Proof of Proposition \ref{prop:trace}} 
\label{app:prop1}

\begin{proof}
Note that the expression inside the brackets of \eqref{eq: joint-loglikelihood} is a univariate function of $D_{i[uv]}$ given $Z_{uv}$. Let $x=D_{i[uv]}$, $a=A_{i[uv]}$, $\mu=Z_{uv}$. Then $\Pi_{i[uv]}=\pi(\mu+x)$, where $\pi(x)\coloneqq1/[1+\exp(-x)]\in(0,1)$. Let $h(x)\coloneqq\log[1-\pi(x)]$. Then 
\begin{equation}
h^{\prime}(x)=\dfrac{-\pi^{\prime}(x)}{1-\pi(x)}=\dfrac{-\pi(x)(1-\pi(x))}{1-\pi(x)}=-\pi(x).\label{eq:1d_of_h}
\end{equation}
Consider $\mu$ as known and the expression inside the brackets of \eqref{eq: joint-loglikelihood} is defined as 
\[
f(x)\coloneqq a\mu+ax+h(\mu+x).
\]
Next we will show that given $\mu$,
\begin{equation}
\underset{x}{\argmax\ }f(x)=\underset{x}{\argmax\ }[a-\pi(\mu)]x.\label{eq:lemma}
\end{equation}
According to \eqref{eq:1d_of_h}, the first derivative of $f(x)$ becomes
\begin{eqnarray*}
f^{\prime}(x) & = & a-\pi(\mu+x).
\end{eqnarray*}
Note that $a\in\{0,1\}$ since $a$ is the realization of a binary random variable. 

(i) $a=1$. $f^{\prime}(x)=1-\pi(\mu+x)>0$ indicating that $f(x)$ is maximized at $x=+\infty$ which also maximizes $[a-\pi(\mu)]x$.

(ii) $a=0$. $f^{\prime}(x)=-\pi(\mu+x)<0$ indicating that $f(x)$ is maximized at $x=-\infty$ which also maximizes $[a-\pi(\mu)]x$.

Then \eqref{eq:lemma} is verified based on (i) and (ii) and the optimal $\{D_{i}:i=1,\dots,n\}$ maximizing \eqref{eq: joint-loglikelihood} given $Z$ can be written as 

\begin{align*}
\underset{\{D_{i}\}}{\argmax\ } & \sum_{i=1}^{n}\sum_{u=1}^{V}\sum_{v<u}\left[A_{i[uv]}(Z_{uv}+D_{i[uv]})+\log(1-\Pi_{i[uv]})\right]\\
= & \underset{\{D_{i}\}}{\argmax\ }\sum_{i=1}^{n}\sum_{u=1}^{V}\sum_{v<u}\left[A_{i[uv]}-\pi(Z_{uv})\right]D_{i[uv]}\\
= & \underset{\{D_{i}\}}{\argmax\ }\sum_{i=1}^{n}\dfrac{1}{2}\tr \left([A_{i}-\pi(Z)]D_{i}\right).
\end{align*}
The last line follows because $\pi(Z$) and each $A_{i}$ are symmetric matrices and their diagonal elements are set at 0.
\end{proof}

\subsection{Proof of Proposition \ref{prop:prop2}}
\label{app:prop2}

\begin{proof}
It suffices to prove \eqref{eq:prop_sup} as \eqref{eq:prop_inf} follows by replacing $B$ with $-B$ noting that $\sigma_{j}(-B)=-\sigma_{V-j+1}(B)$, $j=1,\dots,V$.

We do induction on dimension $k$ and first verify the $k=1$ case. By Rayleigh-Ritz Theorem \citep{parlett1998sym_eigen}, for any unit vector $\boldsymbol{u}\in\mathbb{R}^{V}$ we have
\[
\underset{\boldsymbol{u}^{\top}\boldsymbol{u}=1}{\max}\boldsymbol{u}^{\top}B\boldsymbol{u} = \sigma_{1}(B).
\]
Since $c_{1}>0$, then $\underset{\boldsymbol{u}^{\top}\boldsymbol{u}=1}{\max}c_{1}\boldsymbol{u}^{\top}B\boldsymbol{u} = c_{1}\sigma_{1}(B)$. So \eqref{eq:prop_sup} holds for $k=1$.

Assume \eqref{eq:prop_sup} holds for $k=j-1$. We now show that \eqref{eq:prop_sup} also holds for $k=j$.

Let $\boldsymbol{u}_{1},\dots,\boldsymbol{u}_{j}$ be an orthonormal basis of a $j$-dimensional subspace $U$ in $\mathbb{R}^{V}$. We define a scaled partial trace to represent the objective function in \eqref{eq:prop_sup} for notational simplicity in the rest proof :  
\[
\ptr (B\mid U,c_{1:j})\coloneqq\sum_{i=1}^{j}c_{i}\boldsymbol{u}_{i}^{\top}B\boldsymbol{u}_{i}.
\]
Then 
\[
\underset{\mbox{dim}(U)=j}{\max}\ptr (B\mid U,c_{1:j}) = \underset{\boldsymbol{u}_{1},\dots,\boldsymbol{u}_{j}}{\max}\sum_{i=1}^{j}c_{i}\boldsymbol{u}_{i}^{\top}B\boldsymbol{u}_{i}
\]
for any orthonormal set $\{\boldsymbol{u}_{1},\dots,\boldsymbol{u}_{j}\}$ in $\mathbb{R}^{V}$.

According to Courant-Fischer Theorem \citep{parlett1998sym_eigen}, 
\begin{equation}
\sigma_{j}(B)=\underset{\mbox{dim}(U)=j}{\max}\underset{\begin{array}{c}
\boldsymbol{u}\in U:\boldsymbol{u}^{\top}\boldsymbol{u}=1\end{array}}{\min}\boldsymbol{u}^{\top}B\boldsymbol{u}.\label{eq:courant-fischer}
\end{equation}
Then for every $j$-dimensional subspace $U$ of $\mathbb{R}^{V}$ and any orthonormal basis of $U$, $\boldsymbol{u}_{1},\dots,\boldsymbol{u}_{j}$, there is some $\boldsymbol{u}_{m}$ ($m\in\{1,\dots,j\}$) such that $\boldsymbol{u}_{m}^{\top}B\boldsymbol{u}_{m}\leq\sigma_{j}(B)$. Since $c_{j}>0$, then
\begin{equation}
c_{j}\boldsymbol{u}_{m}^{\top}B\boldsymbol{u}_{m}\leq c_{j}\sigma_{j}(B).\label{eq:piece1}
\end{equation}
The remaining vectors $\{\boldsymbol{u}_{i}:i\neq m\}$ is also an orthonormal basis of a $(j-1)$-dimensional subspace $\tilde{U}$. By induction,
\begin{equation}
\ptr (B\mid\tilde{U},c_{1:(j-1)}) \leq\sum_{i=1}^{j-1}c_{i}\sigma_{i}(B).\label{eq:piece2}
\end{equation}
Adding the two inequalities \eqref{eq:piece1} and \eqref{eq:piece2}, we have
\[
\ptr (B\mid U,c_{1:j})\leq c_{1}\sigma_{1}(B)+\cdots+c_{j}\sigma_{j}(B)
\]
for any $j$-dimensional subspace $U$. Therefore 
\[\underset{\mbox{dim}(U)=j}{\max}\ptr (B\mid U,c_{1:j})\leq c_{1}\sigma_{1}(B)+\cdots+c_{j}\sigma_{j}(B).\]

On the other hand, by selecting $U$ to be the span of the first $j$ orthonormal eigenvectors of $B$, we obtain the reverse inequality
\[
\underset{\mbox{dim}(U)=j}{\max}\ptr (B\mid U,c_{1:j})\geq c_{1}\sigma_{1}(B)+\cdots+c_{j}\sigma_{j}(B).
\]
\end{proof}

\subsection{Inference of $Q$ in the joint embedding model $D_{i}=Q\Lambda_{i}Q^{\top}$}
\label{inference:jem}

We are going to solve $\boldsymbol{q}_{1},...,\boldsymbol{q}_{K}$ sequentially, where $\boldsymbol{q}_{k}$ is the $k$th column of $Q$.  Let $\mbox{eval}_{1}(W)$ denote the largest eigenvalue of $W$. Suppose $s_{1}=\underset{k}{\argmax}\ \mbox{eval}_{1}(W_{k})$. Then set $\boldsymbol{q}_{s_{1}}=\mbox{evec}_{1}(W_{s_{1}})$. To decide the next $\boldsymbol{q}_{k}$ to update, let $U$ be a $V\times(V-1)$ matrix comprising of a set of orthonormal basis of the space orthogonal to $\boldsymbol{q}_{s_{1}}$. For $k\neq s_{1}$, we know $\boldsymbol{q}_{k}\in\mbox{span}(U)$ and hence assume $\boldsymbol{q}_{k}=U\boldsymbol{a}_{k}$ for some vector $\boldsymbol{a}_{k}\in\mathbb{R}^{V-1}$. $\boldsymbol{q}_{k}^{\top}\boldsymbol{q}_{k}=1$ implies that $\boldsymbol{a}_{k}^{\top}U^{\top}U\boldsymbol{a}_{k}=\boldsymbol{a}_{k}^{\top}\boldsymbol{a}_{k}=1$. So $\boldsymbol{a}_{k}$ is of unit length. Then the optimization problem $\underset{\boldsymbol{q}_{k}}{\max}\{\boldsymbol{q}_{k}^{\top}W_{k}\boldsymbol{q}_{k}:\boldsymbol{q}_{k}^{\top}\boldsymbol{q}_{k}=1;\boldsymbol{q}_{k}^{\top}\boldsymbol{q}_{s_{1}}=0\}$ transforms to the optimization 
\begin{eqnarray}
\underset{\boldsymbol{a}_{k}\in\mathbb{R}^{V-1}}{\max} & \boldsymbol{a}_{k}^{\top}U^{\top}W_{k}U\boldsymbol{a}_{k}\label{eq:V-1 space}\\
\mbox{s.t.} & \boldsymbol{a}_{k}^{\top}\boldsymbol{a}_{k}=1\nonumber 
\end{eqnarray}
By Rayleigh-Ritz Theorem, the solution of $\boldsymbol{a}_{k}$ is $\mbox{evec}_{1}(U^{\top}W_{k}U)$ and 
$$\max\,\{\boldsymbol{q}_{k}^{\top}W_{k}\boldsymbol{q}_{k}:\boldsymbol{q}_{k}^{\top}\boldsymbol{q}_{k}=1,\boldsymbol{q}_{k}^{\top}\boldsymbol{q}_{s_{1}}=0\}=\text{\mbox{eval}}_{1}(U^{\top}W_{k}U).$$ Note that the eigenvalues of a symmetric matrix are invariant to orthogonal transformation, i.e. $\mbox{eval}_{1}(U^{\top}W_{k}U)=\mbox{eval}_{1}(R^{\top}U^{\top}W_{k}UR)$ for any orthogonal matrix $R$. Hence U in \eqref{eq:V-1 space} can be an arbitrary orthonormal basis of the subspace $\boldsymbol{q}_{k}^{\bot}$. Suppose $s_{2}=\underset{k\neq s_{1}}{\argmax}\ \mbox{eval}_{1}(U^{\top}W_{k}U)$. Then set $\boldsymbol{q}_{s_{2}}=U\boldsymbol{a}_{s_{2}}$ where $\boldsymbol{a}_{s_{2}}=\mbox{evec}_{1}(U^{\top}W_{s_{2}}U)$. Repeat the above process and we can obtain the other $\boldsymbol{q}_{k}$'s. We summarize the procedure to solve $Q$ given $Z$ and $\{\lambda_{ik}\}$ in Algorithm \ref{alg:Inference_Q}. 

\begin{algorithm}[hb]
Let $W_{k}=\sum_{i=1}^{n}\lambda_{ik}[A_{i}-\pi(Z)],\ k=1,\dots,K.$ \\
Find $s_{1}=\underset{k}{\argmax}\ \mbox{eval}_{1}(W_{k})$ and set $\boldsymbol{q}_{s_{1}}=\mbox{evec}_{1}(W_{s_{1}})$. 

\For {$k=2:K$} {
	find a set of orthogonal basis $U$ of the subspace $\mbox{span}(\boldsymbol{q}_{s_{1}},\dots,\boldsymbol{q}_{s_{(k-1)}})^{\bot}$ \;
	find $s_{k}=\underset{k\in\{1,\dots K\}\setminus\{s_{1},\dots,s_{(k-1)}\}}{\argmax}\ \mbox{eval}_{1}(U^{\top}W_{k}U)$ \;
	set $\boldsymbol{q}_{s_{k}}=U\boldsymbol{a}_{s_{k}}$ where $\boldsymbol{a}_{s_{k}}=\mbox{evec}_{1}(U^{\top}W_{s_{k}}U)$ \;
}
\KwOut{$Q=(\boldsymbol{q}_{1},\dots,\boldsymbol{q}_{K})$.}

\caption{Inference of $Q$ in the joint embedding model $D_{i}=Q\Lambda_{i}Q^{\top}$.\label{alg:Inference_Q}}
\end{algorithm}

\begin{supplement}
\sname{Supplement A}\label{suppA}
\stitle{Code and Data}
\slink[url]{https://github.com/wangronglu/CISE-algorithm}
\sdescription{The R and Matlab codes for CISE algorithm and the HCP data can be found in the link above.}
\end{supplement}

\section*{Acknowledgements}
We would like to thank David Choi for insightful comments, and Peter Hoff for useful comments on method comparisons. This work was partially supported by the grant N00014-14-1-0245 of the United States Office of Naval Research (ONR) and grant W911NF-16-1-0544 of the Army Research Institute (ARI).

\bibliographystyle{imsart-nameyear}
\bibliography{logistic_RDPG}

\begin{thebibliography}{31}

\bibitem[\protect\citeauthoryear{Barch et~al.}{2013}]{barch2013function}
\begin{barticle}[author]
\bauthor{\bsnm{Barch},~\bfnm{Deanna~M}\binits{D.~M.}},
  \bauthor{\bsnm{Burgess},~\bfnm{Gregory~C}\binits{G.~C.}},
  \bauthor{\bsnm{Harms},~\bfnm{Michael~P}\binits{M.~P.}},
  \bauthor{\bsnm{Petersen},~\bfnm{Steven~E}\binits{S.~E.}},
  \bauthor{\bsnm{Schlaggar},~\bfnm{Bradley~L}\binits{B.~L.}},
  \bauthor{\bsnm{Corbetta},~\bfnm{Maurizio}\binits{M.}},
  \bauthor{\bsnm{Glasser},~\bfnm{Matthew~F}\binits{M.~F.}},
  \bauthor{\bsnm{Curtiss},~\bfnm{Sandra}\binits{S.}},
  \bauthor{\bsnm{Dixit},~\bfnm{Sachin}\binits{S.}},
  \bauthor{\bsnm{Feldt},~\bfnm{Cindy}\binits{C.}} \betal{et~al.}
(\byear{2013}).
\btitle{Function in the human connectome: task-fMRI and individual differences
  in behavior}.
\bjournal{Neuroimage}
\bvolume{80}
\bpages{169--189}.
\end{barticle}
\endbibitem

\bibitem[\protect\citeauthoryear{Benjamini and
  Hochberg}{1995}]{benjamini1995controlling}
\begin{barticle}[author]
\bauthor{\bsnm{Benjamini},~\bfnm{Yoav}\binits{Y.}} \AND
  \bauthor{\bsnm{Hochberg},~\bfnm{Yosef}\binits{Y.}}
(\byear{1995}).
\btitle{Controlling the false discovery rate: a practical and powerful approach
  to multiple testing}.
\bjournal{Journal of the Royal Statistical Society. Series B (Methodological)}
\bvolume{57}
\bpages{289--300}.
\end{barticle}
\endbibitem

\bibitem[\protect\citeauthoryear{Desikan et~al.}{2006}]{Desikan2006968}
\begin{barticle}[author]
\bauthor{\bsnm{Desikan},~\bfnm{Rahul~S.}\binits{R.~S.}},
  \bauthor{\bsnm{S{\'e}gonne},~\bfnm{Florent}\binits{F.}},
  \bauthor{\bsnm{Fischl},~\bfnm{Bruce}\binits{B.}},
  \bauthor{\bsnm{Quinn},~\bfnm{Brian~T.}\binits{B.~T.}},
  \bauthor{\bsnm{Dickerson},~\bfnm{Bradford~C.}\binits{B.~C.}},
  \bauthor{\bsnm{Blacker},~\bfnm{Deborah}\binits{D.}},
  \bauthor{\bsnm{Buckner},~\bfnm{Randy~L.}\binits{R.~L.}},
  \bauthor{\bsnm{Dale},~\bfnm{Anders~M.}\binits{A.~M.}},
  \bauthor{\bsnm{Maguire},~\bfnm{R.~Paul}\binits{R.~P.}},
  \bauthor{\bsnm{Hyman},~\bfnm{Bradley~T.}\binits{B.~T.}},
  \bauthor{\bsnm{Albert},~\bfnm{Marilyn~S.}\binits{M.~S.}} \AND
  \bauthor{\bsnm{Killiany},~\bfnm{Ronald~J.}\binits{R.~J.}}
(\byear{2006}).
\btitle{An automated labeling system for subdividing the human cerebral cortex
  on MRI scans into gyral based regions of interest}.
\bjournal{NeuroImage}
\bvolume{31}
\bpages{968 - 980}.
\bdoi{http://dx.doi.org/10.1016/j.neuroimage.2006.01.021}
\end{barticle}
\endbibitem

\bibitem[\protect\citeauthoryear{Dong et~al.}{2014}]{dong2014clustering}
\begin{barticle}[author]
\bauthor{\bsnm{Dong},~\bfnm{Xiaowen}\binits{X.}},
  \bauthor{\bsnm{Frossard},~\bfnm{Pascal}\binits{P.}},
  \bauthor{\bsnm{Vandergheynst},~\bfnm{Pierre}\binits{P.}} \AND
  \bauthor{\bsnm{Nefedov},~\bfnm{Nikolai}\binits{N.}}
(\byear{2014}).
\btitle{Clustering on multi-layer graphs via subspace analysis on Grassmann
  manifolds}.
\bjournal{IEEE Transactions on Signal Processing}
\bvolume{62}
\bpages{905--918}.
\end{barticle}
\endbibitem

\bibitem[\protect\citeauthoryear{Durante, Dunson and
  Vogelstein}{2017}]{durante2017nonparametric}
\begin{barticle}[author]
\bauthor{\bsnm{Durante},~\bfnm{Daniele}\binits{D.}},
  \bauthor{\bsnm{Dunson},~\bfnm{David~B}\binits{D.~B.}} \AND
  \bauthor{\bsnm{Vogelstein},~\bfnm{Joshua~T}\binits{J.~T.}}
(\byear{2017}).
\btitle{Nonparametric Bayes modeling of populations of networks}.
\bjournal{Journal of the American Statistical Association}
\bvolume{112}
\bpages{1516-1530}.
\end{barticle}
\endbibitem

\bibitem[\protect\citeauthoryear{Firth}{1993}]{firth1993bias}
\begin{barticle}[author]
\bauthor{\bsnm{Firth},~\bfnm{David}\binits{D.}}
(\byear{1993}).
\btitle{Bias reduction of maximum likelihood estimates}.
\bjournal{Biometrika}
\bvolume{80}
\bpages{27--38}.
\end{barticle}
\endbibitem

\bibitem[\protect\citeauthoryear{Friedman, Hastie and
  Tibshirani}{2010}]{Friedman2010aa}
\begin{barticle}[author]
\bauthor{\bsnm{Friedman},~\bfnm{Jerome}\binits{J.}},
  \bauthor{\bsnm{Hastie},~\bfnm{Trevor}\binits{T.}} \AND
  \bauthor{\bsnm{Tibshirani},~\bfnm{Robert}\binits{R.}}
(\byear{2010}).
\btitle{Regularization paths for generalized linear models via coordinate
  descent}.
\bjournal{Journal of Statistical Software}
\bvolume{33}
\bpages{1--22}.
\end{barticle}
\endbibitem

\bibitem[\protect\citeauthoryear{Gelman et~al.}{2008}]{gelman2008glm_prior}
\begin{barticle}[author]
\bauthor{\bsnm{Gelman},~\bfnm{Andrew}\binits{A.}},
  \bauthor{\bsnm{Jakulin},~\bfnm{Aleks}\binits{A.}},
  \bauthor{\bsnm{Pittau},~\bfnm{Maria~Grazia}\binits{M.~G.}} \AND
  \bauthor{\bsnm{Su},~\bfnm{Yu-Sung}\binits{Y.-S.}}
(\byear{2008}).
\btitle{A weakly informative default prior distribution for logistic and other
  regression models}.
\bjournal{The Annals of Applied Statistics}
\bvolume{4}
\bpages{1360--1383}.
\end{barticle}
\endbibitem

\bibitem[\protect\citeauthoryear{Girvan and Newman}{2002}]{girvan2002community}
\begin{barticle}[author]
\bauthor{\bsnm{Girvan},~\bfnm{Michelle}\binits{M.}} \AND
  \bauthor{\bsnm{Newman},~\bfnm{Mark~EJ}\binits{M.~E.}}
(\byear{2002}).
\btitle{Community structure in social and biological networks}.
\bjournal{Proceedings of the National Academy of Sciences}
\bvolume{99}
\bpages{7821--7826}.
\end{barticle}
\endbibitem

\bibitem[\protect\citeauthoryear{Goldenberg
  et~al.}{2010}]{goldenberg2010survey}
\begin{barticle}[author]
\bauthor{\bsnm{Goldenberg},~\bfnm{Anna}\binits{A.}},
  \bauthor{\bsnm{Zheng},~\bfnm{Alice~X}\binits{A.~X.}},
  \bauthor{\bsnm{Fienberg},~\bfnm{Stephen~E}\binits{S.~E.}},
  \bauthor{\bsnm{Airoldi},~\bfnm{Edoardo~M}\binits{E.~M.}} \betal{et~al.}
(\byear{2010}).
\btitle{A survey of statistical network models}.
\bjournal{Foundations and Trends in Machine Learning}
\bvolume{2}
\bpages{129--233}.
\end{barticle}
\endbibitem

\bibitem[\protect\citeauthoryear{Heinze}{2006}]{heinze2006penalize}
\begin{barticle}[author]
\bauthor{\bsnm{Heinze},~\bfnm{Georg}\binits{G.}}
(\byear{2006}).
\btitle{A comparative investigation of methods for logistic regression with
  separated or nearly separated data}.
\bjournal{Statistics In Medicine}
\bvolume{25}
\bpages{4216--4226}.
\end{barticle}
\endbibitem

\bibitem[\protect\citeauthoryear{Hoff}{2008}]{hoff2008eigenmodel}
\begin{binproceedings}[author]
\bauthor{\bsnm{Hoff},~\bfnm{Peter}\binits{P.}}
(\byear{2008}).
\btitle{Modeling homophily and stochastic equivalence in symmetric relational
  data}.
In \bbooktitle{Advances in Neural Information Processing Systems}
\bpages{657--664}.
\bpublisher{Curran Associates, Inc.}
\end{binproceedings}
\endbibitem

\bibitem[\protect\citeauthoryear{Hoff, Raftery and
  Handcock}{2002}]{hoff2002latent}
\begin{barticle}[author]
\bauthor{\bsnm{Hoff},~\bfnm{Peter~D}\binits{P.~D.}},
  \bauthor{\bsnm{Raftery},~\bfnm{Adrian~E}\binits{A.~E.}} \AND
  \bauthor{\bsnm{Handcock},~\bfnm{Mark~S}\binits{M.~S.}}
(\byear{2002}).
\btitle{Latent space approaches to social network analysis}.
\bjournal{Journal of the American Statistical Association}
\bvolume{97}
\bpages{1090--1098}.
\end{barticle}
\endbibitem

\bibitem[\protect\citeauthoryear{Kolda and Bader}{2009}]{kolda2009tensor}
\begin{barticle}[author]
\bauthor{\bsnm{Kolda},~\bfnm{Tamara~G}\binits{T.~G.}} \AND
  \bauthor{\bsnm{Bader},~\bfnm{Brett~W}\binits{B.~W.}}
(\byear{2009}).
\btitle{Tensor decompositions and applications}.
\bjournal{SIAM Review}
\bvolume{51}
\bpages{455--500}.
\end{barticle}
\endbibitem

\bibitem[\protect\citeauthoryear{Kravitz et~al.}{2011}]{kravitz2011new}
\begin{barticle}[author]
\bauthor{\bsnm{Kravitz},~\bfnm{Dwight~J}\binits{D.~J.}},
  \bauthor{\bsnm{Saleem},~\bfnm{Kadharbatcha~S}\binits{K.~S.}},
  \bauthor{\bsnm{Baker},~\bfnm{Chris~I}\binits{C.~I.}} \AND
  \bauthor{\bsnm{Mishkin},~\bfnm{Mortimer}\binits{M.}}
(\byear{2011}).
\btitle{A new neural framework for visuospatial processing}.
\bjournal{Nature Reviews Neuroscience}
\bvolume{12}
\bpages{217--230}.
\end{barticle}
\endbibitem

\bibitem[\protect\citeauthoryear{Lock et~al.}{2013}]{lock2013joint}
\begin{barticle}[author]
\bauthor{\bsnm{Lock},~\bfnm{Eric~F}\binits{E.~F.}},
  \bauthor{\bsnm{Hoadley},~\bfnm{Katherine~A}\binits{K.~A.}},
  \bauthor{\bsnm{Marron},~\bfnm{James~Stephen}\binits{J.~S.}} \AND
  \bauthor{\bsnm{Nobel},~\bfnm{Andrew~B}\binits{A.~B.}}
(\byear{2013}).
\btitle{Joint and individual variation explained (JIVE) for integrated analysis
  of multiple data types}.
\bjournal{The Annals of Applied Statistics}
\bvolume{7}
\bpages{523}.
\end{barticle}
\endbibitem

\bibitem[\protect\citeauthoryear{Minka}{2003}]{minka2003comparison}
\begin{barticle}[author]
\bauthor{\bsnm{Minka},~\bfnm{Thomas~P}\binits{T.~P.}}
(\byear{2003}).
\btitle{A comparison of numerical optimizers for logistic regression}.
\bjournal{Unpublished draft}.
\end{barticle}
\endbibitem

\bibitem[\protect\citeauthoryear{Moore et~al.}{2015}]{moore2015development}
\begin{barticle}[author]
\bauthor{\bsnm{Moore},~\bfnm{Tyler~M}\binits{T.~M.}},
  \bauthor{\bsnm{Scott},~\bfnm{J~Cobb}\binits{J.~C.}},
  \bauthor{\bsnm{Reise},~\bfnm{Steven~P}\binits{S.~P.}},
  \bauthor{\bsnm{Port},~\bfnm{Allison~M}\binits{A.~M.}},
  \bauthor{\bsnm{Jackson},~\bfnm{Chad~T}\binits{C.~T.}},
  \bauthor{\bsnm{Ruparel},~\bfnm{Kosha}\binits{K.}},
  \bauthor{\bsnm{Savitt},~\bfnm{Adam~P}\binits{A.~P.}},
  \bauthor{\bsnm{Gur},~\bfnm{Raquel~E}\binits{R.~E.}} \AND
  \bauthor{\bsnm{Gur},~\bfnm{Ruben~C}\binits{R.~C.}}
(\byear{2015}).
\btitle{Development of an abbreviated form of the Penn Line Orientation Test
  using large samples and computerized adaptive test simulation.}
\bjournal{Psychological Assessment}
\bvolume{27}
\bpages{955}.
\end{barticle}
\endbibitem

\bibitem[\protect\citeauthoryear{Newman}{2010}]{newman2010networks}
\begin{bbook}[author]
\bauthor{\bsnm{Newman},~\bfnm{Mark}\binits{M.}}
(\byear{2010}).
\btitle{Networks: an introduction}.
\bpublisher{Oxford university press}.
\end{bbook}
\endbibitem

\bibitem[\protect\citeauthoryear{O'Connor, M{\'e}dard and
  Feizi}{2015}]{o2015clustering}
\begin{barticle}[author]
\bauthor{\bsnm{O'Connor},~\bfnm{Luke}\binits{L.}},
  \bauthor{\bsnm{M{\'e}dard},~\bfnm{Muriel}\binits{M.}} \AND
  \bauthor{\bsnm{Feizi},~\bfnm{Soheil}\binits{S.}}
(\byear{2015}).
\btitle{Clustering over logistic random dot product graphs}.
\bjournal{Stat}
\bvolume{1050}
\bpages{3}.
\end{barticle}
\endbibitem

\bibitem[\protect\citeauthoryear{Parlett}{1998}]{parlett1998sym_eigen}
\begin{bbook}[author]
\bauthor{\bsnm{Parlett},~\bfnm{Beresford~N}\binits{B.~N.}}
(\byear{1998}).
\btitle{The symmetric eigenvalue problem}.
\bpublisher{SIAM}.
\end{bbook}
\endbibitem

\bibitem[\protect\citeauthoryear{Sussman et~al.}{2012}]{sussman2012consistent}
\begin{barticle}[author]
\bauthor{\bsnm{Sussman},~\bfnm{Daniel~L}\binits{D.~L.}},
  \bauthor{\bsnm{Tang},~\bfnm{Minh}\binits{M.}},
  \bauthor{\bsnm{Fishkind},~\bfnm{Donniell~E}\binits{D.~E.}} \AND
  \bauthor{\bsnm{Priebe},~\bfnm{Carey~E}\binits{C.~E.}}
(\byear{2012}).
\btitle{A consistent adjacency spectral embedding for stochastic blockmodel
  graphs}.
\bjournal{Journal of the American Statistical Association}
\bvolume{107}
\bpages{1119--1128}.
\end{barticle}
\endbibitem

\bibitem[\protect\citeauthoryear{Tang, Lu and
  Dhillon}{2009}]{tang2009clustering}
\begin{binproceedings}[author]
\bauthor{\bsnm{Tang},~\bfnm{Wei}\binits{W.}},
  \bauthor{\bsnm{Lu},~\bfnm{Zhengdong}\binits{Z.}} \AND
  \bauthor{\bsnm{Dhillon},~\bfnm{Inderjit~S}\binits{I.~S.}}
(\byear{2009}).
\btitle{Clustering with multiple graphs}.
In \bbooktitle{IEEE International Conference on Data Mining}
\bpages{1016--1021}.
\end{binproceedings}
\endbibitem

\bibitem[\protect\citeauthoryear{Tang et~al.}{2016}]{tang2016law}
\begin{barticle}[author]
\bauthor{\bsnm{Tang},~\bfnm{Runze}\binits{R.}},
  \bauthor{\bsnm{Ketcha},~\bfnm{Michael}\binits{M.}},
  \bauthor{\bsnm{Vogelstein},~\bfnm{Joshua~T}\binits{J.~T.}},
  \bauthor{\bsnm{Priebe},~\bfnm{Carey~E}\binits{C.~E.}} \AND
  \bauthor{\bsnm{Sussman},~\bfnm{Daniel~L}\binits{D.~L.}}
(\byear{2016}).
\btitle{Law of large graphs}.
\bjournal{arXiv preprint arXiv:1609.01672}.
\end{barticle}
\endbibitem

\bibitem[\protect\citeauthoryear{Tucker}{1966}]{tucker1966some}
\begin{barticle}[author]
\bauthor{\bsnm{Tucker},~\bfnm{Ledyard~R}\binits{L.~R.}}
(\byear{1966}).
\btitle{Some mathematical notes on three-mode factor analysis}.
\bjournal{Psychometrika}
\bvolume{31}
\bpages{279--311}.
\end{barticle}
\endbibitem

\bibitem[\protect\citeauthoryear{Van~Essen et~al.}{2012}]{van2012human}
\begin{barticle}[author]
\bauthor{\bsnm{Van~Essen},~\bfnm{David~C}\binits{D.~C.}},
  \bauthor{\bsnm{Ugurbil},~\bfnm{Kamil}\binits{K.}},
  \bauthor{\bsnm{Auerbach},~\bfnm{E}\binits{E.}},
  \bauthor{\bsnm{Barch},~\bfnm{D}\binits{D.}},
  \bauthor{\bsnm{Behrens},~\bfnm{TEJ}\binits{T.}},
  \bauthor{\bsnm{Bucholz},~\bfnm{R}\binits{R.}},
  \bauthor{\bsnm{Chang},~\bfnm{Acer}\binits{A.}},
  \bauthor{\bsnm{Chen},~\bfnm{Liyong}\binits{L.}},
  \bauthor{\bsnm{Corbetta},~\bfnm{Maurizio}\binits{M.}},
  \bauthor{\bsnm{Curtiss},~\bfnm{Sandra~W}\binits{S.~W.}} \betal{et~al.}
(\byear{2012}).
\btitle{The Human Connectome Project: a data acquisition perspective}.
\bjournal{Neuroimage}
\bvolume{62}
\bpages{2222--2231}.
\end{barticle}
\endbibitem

\bibitem[\protect\citeauthoryear{Woolfe et~al.}{2008}]{woolfe2008fast}
\begin{barticle}[author]
\bauthor{\bsnm{Woolfe},~\bfnm{Franco}\binits{F.}},
  \bauthor{\bsnm{Liberty},~\bfnm{Edo}\binits{E.}},
  \bauthor{\bsnm{Rokhlin},~\bfnm{Vladimir}\binits{V.}} \AND
  \bauthor{\bsnm{Tygert},~\bfnm{Mark}\binits{M.}}
(\byear{2008}).
\btitle{A fast randomized algorithm for the approximation of matrices}.
\bjournal{Applied and Computational Harmonic Analysis}
\bvolume{25}
\bpages{335--366}.
\end{barticle}
\endbibitem

\bibitem[\protect\citeauthoryear{Wu et~al.}{2015}]{wu2015cortical}
\begin{barticle}[author]
\bauthor{\bsnm{Wu},~\bfnm{Xiu}\binits{X.}},
  \bauthor{\bsnm{Lv},~\bfnm{Xiao-Fei}\binits{X.-F.}},
  \bauthor{\bsnm{Zhang},~\bfnm{Yu-Ling}\binits{Y.-L.}},
  \bauthor{\bsnm{Wu},~\bfnm{Hua-Wang}\binits{H.-W.}},
  \bauthor{\bsnm{Cai},~\bfnm{Pei-Qiang}\binits{P.-Q.}},
  \bauthor{\bsnm{Qiu},~\bfnm{Ying-Wei}\binits{Y.-W.}},
  \bauthor{\bsnm{Zhang},~\bfnm{Xue-Lin}\binits{X.-L.}} \AND
  \bauthor{\bsnm{Jiang},~\bfnm{Gui-Hua}\binits{G.-H.}}
(\byear{2015}).
\btitle{Cortical signature of patients with HBV-related cirrhosis without overt
  hepatic encephalopathy: a morphometric analysis}.
\bjournal{Frontiers In Neuroanatomy}
\bvolume{9}
\bpages{82}.
\end{barticle}
\endbibitem

\bibitem[\protect\citeauthoryear{Zeki et~al.}{1991}]{zeki1991direct}
\begin{barticle}[author]
\bauthor{\bsnm{Zeki},~\bfnm{Semir}\binits{S.}},
  \bauthor{\bsnm{Watson},~\bfnm{JD}\binits{J.}},
  \bauthor{\bsnm{Lueck},~\bfnm{CJ}\binits{C.}},
  \bauthor{\bsnm{Friston},~\bfnm{Karl~J}\binits{K.~J.}},
  \bauthor{\bsnm{Kennard},~\bfnm{C}\binits{C.}} \AND
  \bauthor{\bsnm{Frackowiak},~\bfnm{RS}\binits{R.}}
(\byear{1991}).
\btitle{A direct demonstration of functional specialization in human visual
  cortex}.
\bjournal{Journal of Neuroscience}
\bvolume{11}
\bpages{641--649}.
\end{barticle}
\endbibitem

\bibitem[\protect\citeauthoryear{Zhang et~al.}{2018a}]{Zhang2017HCP}
\begin{barticle}[author]
\bauthor{\bsnm{Zhang},~\bfnm{Zhengwu}\binits{Z.}},
  \bauthor{\bsnm{Descoteaux},~\bfnm{Maxime}\binits{M.}},
  \bauthor{\bsnm{Zhang},~\bfnm{Jingwen}\binits{J.}},
  \bauthor{\bsnm{Girard},~\bfnm{Gabriel}\binits{G.}},
  \bauthor{\bsnm{Chamberland},~\bfnm{Maxime}\binits{M.}},
  \bauthor{\bsnm{Dunson},~\bfnm{David}\binits{D.}},
  \bauthor{\bsnm{Srivastava},~\bfnm{Anuj}\binits{A.}} \AND
  \bauthor{\bsnm{Zhu},~\bfnm{Hongtu}\binits{H.}}
(\byear{2018}a).
\btitle{Mapping population-based structural connectomes}.
\bjournal{NeuroImage}
\bvolume{172}
\bpages{130--145}.
\end{barticle}
\endbibitem

\bibitem[\protect\citeauthoryear{Zhang et~al.}{2018b}]{zhang2018relationships}
\begin{barticle}[author]
\bauthor{\bsnm{Zhang},~\bfnm{Zhengwu}\binits{Z.}},
  \bauthor{\bsnm{Allen},~\bfnm{Genevera}\binits{G.}},
  \bauthor{\bsnm{Zhu},~\bfnm{Hongtu}\binits{H.}} \AND
  \bauthor{\bsnm{Dunson},~\bfnm{David}\binits{D.}}
(\byear{2018}b).
\btitle{Relationships between human brain structural connectomes and traits}.
\bjournal{bioRxiv}
\bpages{256933}.
\end{barticle}
\endbibitem

\end{thebibliography}

\end{document}